\documentclass[]{article}
\usepackage{color}
\usepackage[]{graphicx}
\usepackage{authblk}
\usepackage{comment}
\usepackage{ulem}
\usepackage{a4wide} 
\usepackage{amsmath} 

\newcommand{\bm}[1]{\mbox{\boldmath$#1$}}

\title{Effects of multiple cycles on the resistance distance of a strand in a homogeneous polymer network}
\author{Erica Uehara}
\affil{Department of Applied Mathematics and Physics, Graduate School of Informatics, Kyoto University, \\ 
Yoshida-Honmachi, Sakyo-ku, Kyoto City, Kyoto 606-8501, Japan}
\author{Tetsuo Deguchi}
\affil{Department of Physics, Faculty of Core Research, Ochanomizu University, \\ 
2-1-1 Ohtsuka, Bunkyo-ku, Tokyo 112-8610, Japan}
\begin{document}
\maketitle

\begin{abstract}
We show that the resistance distance between a pair of adjacent vertices in a phantom network generated randomly by a Monte-Carlo method depends on the existence of short cycles around it.
Here we assume that phantom networks have no fixed points but their centers of mass are located at a point.
The resistance distance corresponds to the mean-square deviation of the end-to-end vector along the strand connecting the adjacent vertices.
We generate random networks with fixed valency $f$ but different densities of short cycles via a Metropolis method that rewires edges among four vertices chosen randomly.
In the process the cycle rank is conserved. 
However, the densities of short cycles are determined by the rate of randomization $kT$ which appears in the acceptance ratio $\exp(-\Delta U/kT)$ of rewiring.
If a strand has few short cycles around itself, the mean squared deviation of the strand is equal to $2/f$. If it is part of a short cycle, i.e., the network has a short loop which consists of a sequence of strands including the given strand itself, its resistance distance is smaller than $2/f$,
while if it is not included in a cycle but adjacent to cycles, its resistance distance is larger than $2/f$. 
We show it via an electrical circuit analogy of the network.
Moreover, we numerically show that the effect of multiple cycles on the resistance distance is expressed as a linear combination of the effects of isolated single cycles. 
It follows that cycles independently have an effect on the fluctuation properties of a strand in a polymer network.
\end{abstract}

\tableofcontents


\section{Introduction}

Polymer architecture, i.e., the chemical connectivity of polymer strands, characterizes various properties of polymeric materials. Important statistical quantities, e.g., the mean square radius of gyration, the hydrodynamic radius and the scattering intensity of a polymer chain in solution, depend on its architecture such as a linear, ring, branched, etc. \cite{jacs2000oike,macro2007takano,jacs2011sugai,uehara2016jchemphys,wiley2002teraoka}. 
Dynamical quantities also depend on the architecture.  For instance, the viscosity of melt of ring polymers is smaller than that of linear polymers \cite{macro2015doi1}, while the viscosity of melt of lasso polymers is larger \cite{macro2015doi2}.
Simulation shows that diagrams of phase separation depend on the architecture of block copolymers \cite{macro2024tomiyoshi}.
Thus, it is expected that key properties of gels and rubbers should also depend on the architecture. 
However, the architecture of a bulk polymer network is quite complicated and has hardly been specified. 
In this century, Tetra-PEG gels have been successfully synthesized, which are polymeric gels of  highly homogeneous architecture \cite{macro2008sakai}. 
Its synthesis has motivated modern studies on the relationship between architecture and properties of polymeric materials \cite{polyj2013sugimura}, \cite{aipad2018toda}.

The fluctuation of a strand $\overline{\Delta {\bm r}^2}=\overline{({\bm r} - \bar{{\bm r}})^2}$ is given by the resistance distance between the ends of the strand, where ${\bm r}$ is the end-to-end vector between a pair of crosslinks connected with the strand and $\bar{\bm r}$ is its average. 
The resistance distance between a pair of vertices is defined by the resistance between the vertices when each strand in the network is replaced by a unit resistor and current flows into the network from one of the vertices to the other vertex \cite{jmathchem1993klein}. 
We call the resistance distance between a pair of vertices connected with an edge the {\it resistance distance along the edge}. 
Various properties of polymeric materials are relevant to the fluctuation of strands in the polymer network, i.e., the resistance distance along the edge corresponding to the strand. 
Higgs and Ball \cite{jphysf1988higgs} consider a network of a grounded circuit and the fluctuation of a strand  while Cantarella et al.\cite{jphysA2022cantarella} do not require that a circuit is grounded. 
Grounding a circuit corresponds to considering a polymer network having some fixed crosslinks, i.e., the positions of the fixed crosslinks are constant and do not fluctuate. 
De Gennes also compares the macroscopic conductance of a resistor network with the elastic modulus of a gel\cite{jphysique1976deGennes}.

The classical theory on statistical properties of a polymer network has been developed since 1940s \cite{jchemphys1943james}, \cite{jchemphys1947james}, while networks have vertices fixed to the background.
In 1970s Eichinger, Graessley and Flory argued that $\overline{\Delta {\bm r}^2}=2/f$ for a network of constant valency $f$ but the architecture of the network is specified to a tree, which means that the network has no cyclic structures \cite{macro1972Eichinger}  
\cite{macro1975Graessley} \cite{prsla1976flory}. 
Kloczkowski et al. proved it for a complete $f$-ary tree with infinite levels whose boundary sites are fixed to the background \cite{macro1989kloczkowski}. 
Here we remark that they fixed the network to the background in order to avoid the divergence of the partition function.
Moreover, it is nontrivial to calculate statistical properties of a polymer network with cycles, which are local structures of strands connected together in a circuit path.
A cycle leads to constraint such that the sum of the end-to-end vectors along the strands of the cycle vanishes \cite{arxiv2020cantarellauehara}, \cite{arxiv2022cantarellauehara}.
Thus, the conventional approach to the statistical properties of a polymer network has assumptions that the network is acyclic and has fixed points \cite{macro1989kloczkowski},\cite{ox2003rubinstein},\cite{jphysa2015grosberg}.  
However, there are cycles in most of polymer networks in reality. 
Recently, the phantom network theory has been improved so that the network architecture includes a 'defect' \cite{science2016zhong}, \cite{macro2018lang}, \cite{macro2019panyukov}, \cite{macro2019lang}, \cite{macro2019Lin}.
A defect is a local structure such as a cycle, dangling strand or loop strand. 
The researches of Zhong et al. and Lin et al. have shown that a defect decreases the elasticity of the polymer network although a cycle does not change the sum of fluctuations of all strands in a network with a single cycle \cite{science2016zhong},\cite{macro2019Lin}.

While the classical and improved phantom network theories assume that the network is almost acyclic,
the coarse-grained simulations show the mechanical properties of a polymer network with many cycles:
the elasticity of lattice networks with several functionalities was derived by using MD simulation \cite{aipad2018toda}\cite{softmatter2022hagita}. 
The elasticity of lattice networks with defects, percolation networks, was simulated \cite{jchemphys2015nishi}.
Several simulations show the properties of the randomly crosslinked networks \cite{macro2019gusev}, \cite{patterns2020amamoto}, \cite{polymers2021munoz}:
The mechanical properties of the randomly crosslinked networks depend on its modified centralities \cite{patterns2020amamoto}.
The relationship of mechanical properties with heterogeneity, where crosslinks are located randomly or regularly, the number of monomers of strands are same or distributed and the functionalities of the crosslinks are same or distributed \cite{polymers2021munoz}.
The dynamical simulation often has a long computational time to obtain an equilibrium conformation of the networks.

In this research, we investigate the resistance distance along an edge of a random network of fixed valency $f$ generated by a Monte-Carlo method, which has many cycles including both very long cycles (e.g., Hamilton cycles) and short cycles (triangles)  and does not have fixed vertices. By the method we derive  various networks with fixed cycle rank such as lattice-type networks and random regular graph networks which have almost no short cycles and hence are locally close to tree networks. 
We consider the generated network as a Gaussian network where the edges of the network are given by Gaussian chains.
The fluctuation of the end-to-end vector of a Gaussian strand $\bar{r}^2$ is given by the resistance distance along an edge and derived from the Kirchhoff matrix of the network exactly \cite{jphysA2022cantarella}.
We numerically show the effect of multiple cycles on the resistance distance based on our new approach \cite{arxiv2020cantarellauehara}, \cite{arxiv2022cantarellauehara}, \cite{springer2022cantarella}, which does not need fixed boundaries. We instantaneously obtain the statistics of equilibrium conformations of the network without relaxation process.
We also analytically derive the effect of a single cycle on the resistance distance.
The resistance distance along an edge clearly depends on the number of cycles around the edge: 
If the edge is included in the cycle, i.e., the cycle consists of the edge and several edges, the resistance distance along the edge becomes smaller than $2/f$.
And if the edge is not included in the cycle but next to the cycle, the resistance distance along the edge becomes larger than $2/f$.
The mean resistance distance along an edge over all edges in a random $f$-ary network is equal to $2/f$ and independent to the density of cycles.
It is exactly derived from Foster's theorem.

We generate random networks with constant functionality $f$ and the number of vertices $N$ via a Metropolis method.
Generated networks are {\it simple}: 
they have neither loop edges nor multi-edges, which are referred as primary and secondary loops in the improved phantom network theory, respectively \cite{science2016zhong}.
The networks have multiple cycles, while the classical and improved phantom network theories consider almost acyclic, i.e., almost tree-like networks with only one cycle.
Moreover, we can control the density of cycles by changing the acceptance ratio during Monte-Carlo steps.
We randomized a network by rewiring edges with respect to the acceptance ratio $\exp[-\Delta U/kT]$.
McKay uses rewiring techniques to generate random graphs with fixed functionality, where every pair of vertices has an edge with the constant possibility $f/N$ \cite{algo1990mckay}.
We rewires edges in a network so that a pair of geometrically close vertices has an edge with the possibility higher than $f/N$.
Munoz rewires a network \cite{polymers2021munoz}, \cite{macro2022munoz}, where a pair of close crosslinks is connected with a strand so that the total square length of strands is minimal.
Our networks have different expected square lengths of strands which are given by the different rates of randomization $kT$.

The contents of the paper are given in the following: 
In section 2, we provide the method to randomize a given network with a Metropolis algorithm.
It conserves the functionality of the vertices, the connectivity and simplicity of the given network.
In section 3, we show the universality of the mean resistance distance $\overline{\Delta r^2}=f/2$ by using Foster's theorem \cite{ccacaa2002klein} and the discrete distribution of resistance distances.
We also numerically show that the linear dependence of resistance distance along an edge on the number of cycles around the edge.
The effect of multiple cycles is almost equal to the sum of the effects of a single cycle.
In section 4, we give a numerical result of the cycle densities.
As $kT$ becomes smaller, the number of short cycles becomes larger.


\section{Method}

\begin{figure}[htbp]
\begin{center}
\includegraphics[width=6cm]{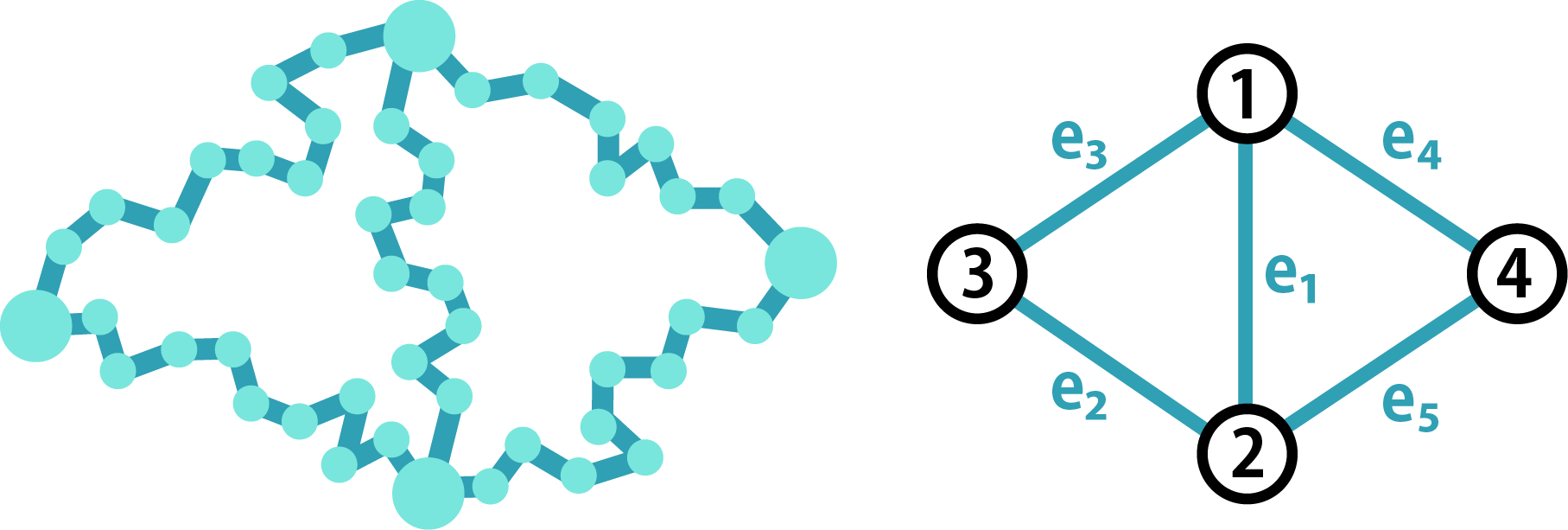}
\caption{Polymer network of $5$ strands and its graph representation.}
\label{fig_PolymerGraph}
\end{center}
\end{figure}

We express the architecture of a polymer network by a graph.
A graph consists of a set of vertices also called nodes and a set of edges also called links.
Each edge has a pair of vertices at its ends.
We illustrate a network with polymer strands consisting of several monomers in the left picture of the figure \ref{fig_PolymerGraph}.
The polymer strands are replaced by the edges of the graph in the right diagram of the figure \ref{fig_PolymerGraph}, and the crosslinks are replaced by the vertices.
Here the configurations of individual monomers in a polymer strand is not considered.

A path is a sequence of edges and vertices placed alternately, where the $(i-1)$-th and $(i+1)$-th vertices are connected by the $i$-th edge.
For example, $(1,e_4,4,e_5,2)$ or simply $(1-4-2)$ is a path in the graph of figure \ref{fig_PolymerGraph}.
A cycle is a path whose initial and last vertices are identical and in this research other vertices are distinct each other.
We call a cycle that has $k$ edges and $k$ vertices a $k$-cycle.
The path $(1-2-3-1)$ in the graph of figure \ref{fig_PolymerGraph} is a $3$-cycle.
The degree of a vertex, also called the functionality or the valency of the vertex, is the number of edges which are incident to the vertex.
An $f$-regular graph is a graph where the degrees of all vertices are equal to $f$.
The cycle rank of an $f$-regular graph of $N$ vertices is $Nf/2-N+1$, the regular graph has many cycles when $N$ is large.

\subsection{Randomization of a regular graph}

We generate regular graphs with different rates of randomization for different functionalities such as $f=3$ or $4$.
Here the rate of randomization is the constant $kT$ in the acceptance ratio $\exp(-\Delta U/kT)$ during the randomization process of a graph, which is illustrated as follows. 

In this paper, we mainly consider the periodic cubic lattice and the template coordinates are given by the coordinates of the $3$-dimensional points on the periodic cubic lattice.

\begin{figure}[htbp]
\includegraphics[width=10cm]{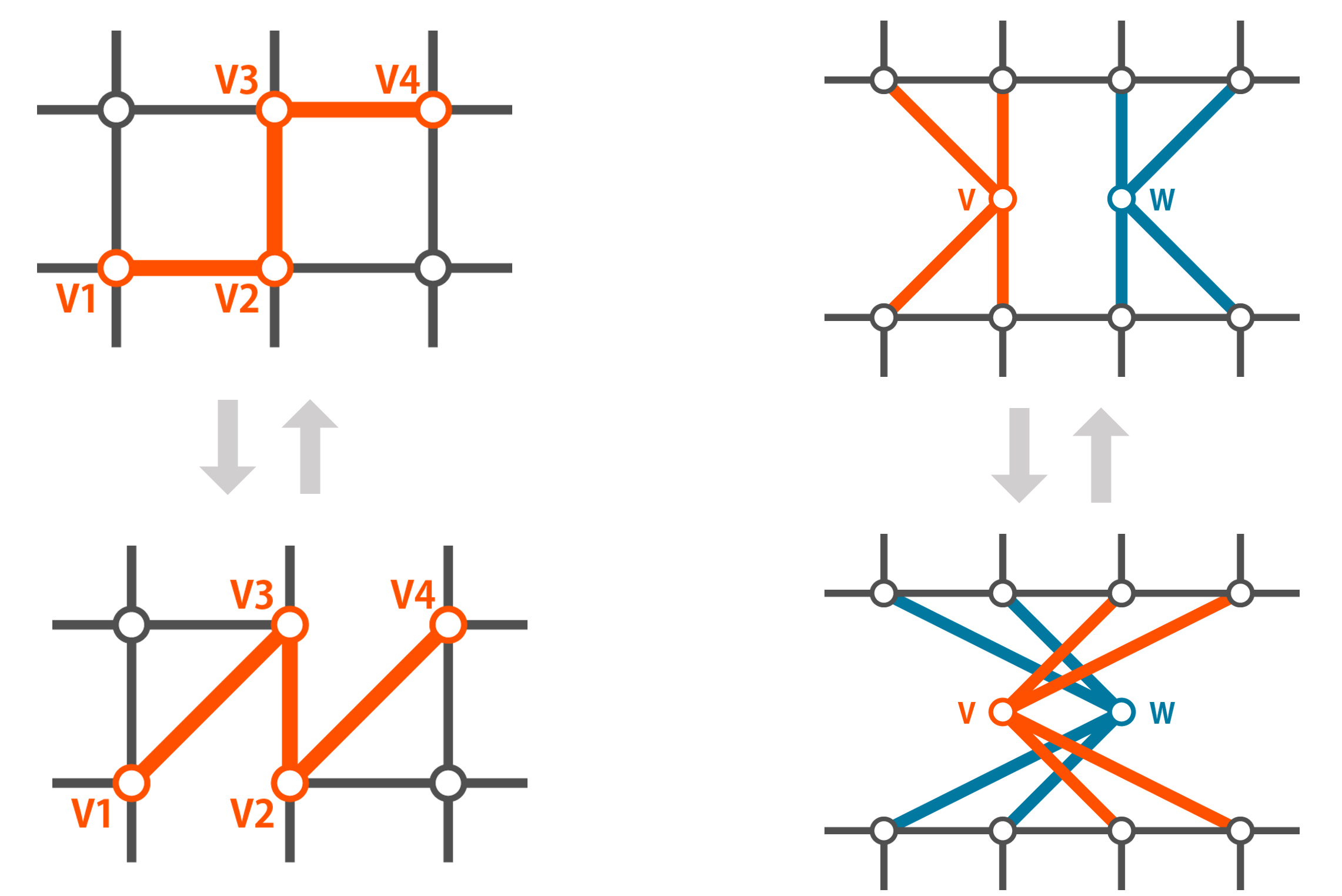}
\caption{Two types of switching procedures. The left diagrams: edge switching. It rewires the edges among $V_1, V_2, V_3$ and $V_4$. The right diagrams: vertex switching. Vertices $V$ and $W$ exchange their incident vertices.}
\label{fig_Switch}
\end{figure}

\begin{enumerate}
\item Assume an initial graph that is given by an arbitrary regular graph.
\item Assign coordinates for all vertices in the initial graph. 
Here we call the coordinates of the vertices in the initial graph the template coordinates.
The given template coordinates are not changed during the randomization. 
\item Assume each edge has an energy with respect to its length. Here we calculate the edge energy as a harmonic spring: 
when a vertex $V$ and a vertex $W$ are connected by an edge and their coordinates are given by $x(V)$ and $x(W)$, respectively, the energy of the edge is given by $\frac{1}{2}(x(V)-x(W))^2$. 
\item Switch edges or vertices randomly as follows. 
(1) Edge switching: We remove a pair of edges and add a new pair of edges in which each edge connects two distinct vertices among the four vertices: We randomly choose a path $(V_1-V_2-V_3-V_4)$ consisting of four vertices in the graph (see the left diagrams of the figure \ref{fig_Switch}). 
Here the path should be self-avoiding, i.e., if $j\neq k$, $V_j \ne V_k$ for $j, k=1,2, \cdots, 4$, and have no edges between the $V_1$ and $V_3$ or between $V_2$ and $V_4$. 
We remove the first edge $(V_1-V_2)$ and the third edge $(V_3-V_4)$. 
Then we add new edges $(V_1-V_3)$ and $(V_2-V_4)$. 
(2) Vertex switching: We exchange a pair of vertices $V$ and $W$, which are randomly chosen in the graph (see the right diagrams of the figure \ref{fig_Switch}): 
Vertex $V$ is connected to vertices $V_1, V_2, V_3, ..., V_f$ and vertex $W$ to vertices $W_1, ...,W_f$. After the switching, $V$ is connected to $W_1, ..., W_f$ and $W$ to $V_1, ..., V_f$. 
Remark that vertex switching generates an isomorphic graph of the given graph, i.e., it swaps only the vertex names and doesn't change the effective structure of the graph. 
However, it changes the edge energies.
\item Calculate the energy difference $\Delta U$ after switching. 
\item If $\Delta U<0$, accept switching. If $\Delta U>0$, generate a uniform random number $\alpha\in[0,1]$. If $\alpha < \exp(-\Delta U/kT)$, accept switching. If not, reject it.
\item Repeat (2-5).
\end{enumerate}

\begin{figure}[htbp]
\begin{center}
\includegraphics[width=7.5cm]{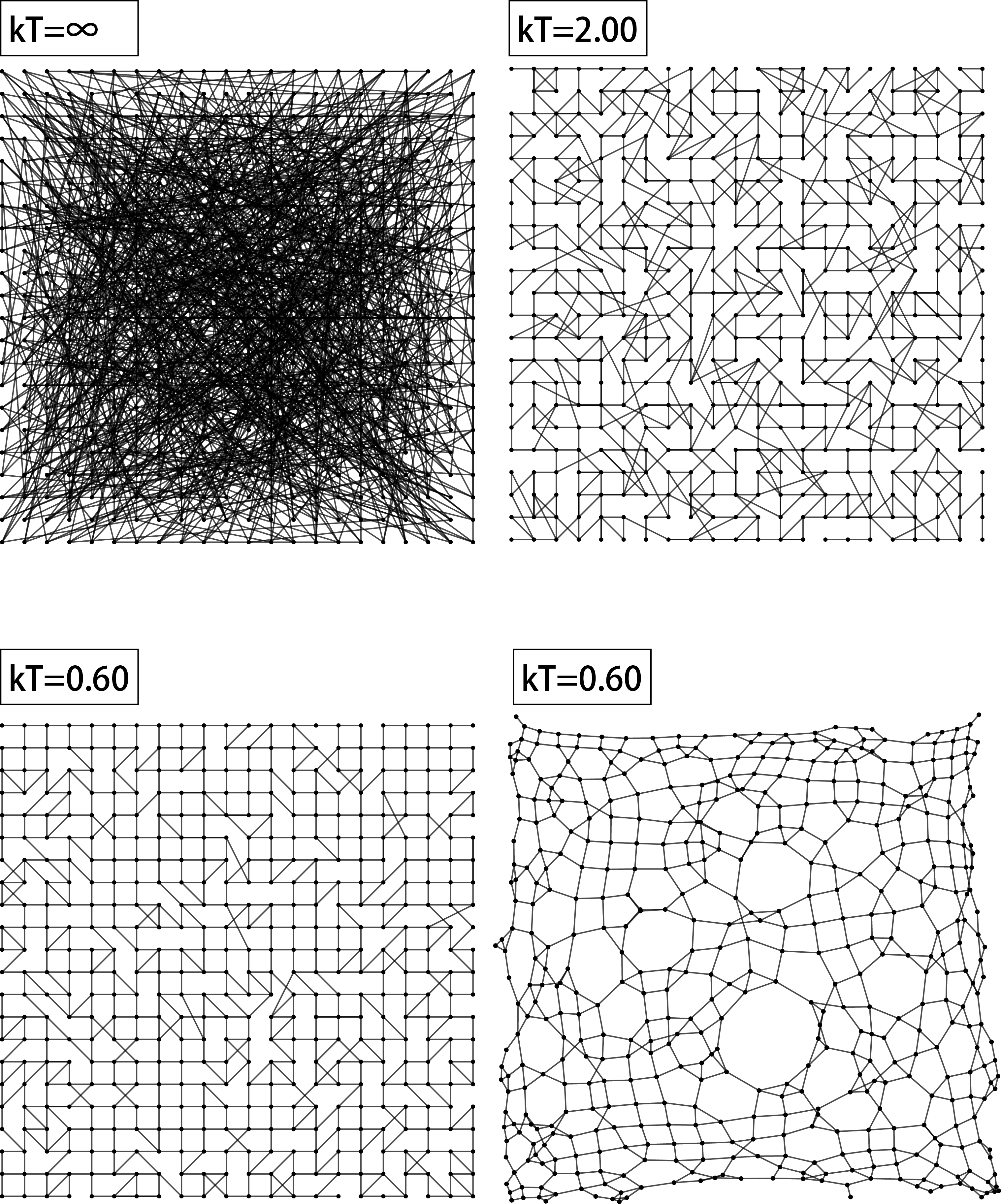}
\caption{4-regular graphs of $484$ vertices. The left upper panel: A 4-regular graph is embedded in the periodic 2-dimensional $22\times 22$ cubic lattice \textcolor{red}{(template coordinates)}. The right upper panel: A 4-regular graph after $40,000,000$ edge switches and $40,000,000$ vertex switches with the rate of randomization $kT=2.00$. We removed the edges across a periodic boundary for visualization. The left bottom panel: A 4-regular graph after $80,000,000$ edge switches and $80,000,000$ vertex switches with the rate of randomization $kT=0.60$. We removed the edges across a periodic boundary for visualization.  The right bottom panel: the same graph to the graph showed in the left bottom panel. The embeddings were given by Methematica12.1.}
\label{fig_graph_2D}
\end{center}
\end{figure}

The randomization process generates various types of regular graphs that include the lattice and the regular graphs illustrated in \cite{algo1990mckay}.
If the rate of randomization $kT$ is equal to an infinity, i.e., we accept all switching procedures, every pair of vertices has an edge with the same probability $f/N$.
If $kT$ is finite, edges tend to be shorter after switching (See the right upper panel of figure \ref{fig_graph_2D}).
Thus the probability of a pair of vertices having an edge depends on the geometrical distance between the vertices.
As $kT$ approaches $0$, the graph approaches a lattice (See the lower panels of figure \ref{fig_graph_2D}).


We have several remarks on this randomization process:
\begin{enumerate}
\item If an initial graph is a connected graph, i.e., every pair of vertices has a path between them, 
all graphs generated from it in the randomization process are connected graphs.

Proof. (1) A pair of vertices $V$ and $W$ has a path including a removed edge like $(V-...-V_1-V_2-V_3-V_4-...-W)$, edge switch removes the path and gives the new path like $(V-...-V_1-V_3-V_4-V_2-...-W)$. Thus it does not break the connectivity of the graph. 
(2) Vertex switches does not change the structure of the graph. 
\item If an initial graph is a simple graph, i.e. it has neither loop edges nor multi-edges, all randomized graphs are also simple.
\item Every vertex does not change its functionality during the randomization procedure.
\item The randomization process reserves the cycle rank of the graph. However, the number of all cycles in the graph can be changed by the process because the cycle rank is the number of fundamental cycles, which is a set of cycles that generates every cycle in a graph by a linear combination of fundamental cycles. Especially, the number of short cycles (3,4,5,6-cycles) increases as $kT$ decreases.
\end{enumerate}

\subsection{Resistance distance}
The graph Laplacian $L$ also known as a Kirchhoff matrix is a square matrix with integer elements.
The $i$-th diagonal element $L_{ii}$ is given by the functionality of the $i$-th vertex
and off-diagonal element $L_{ij}$ is $-1$ if the $i$-th and $j$-th vertices are connected by an edge or $0$ if they are not connected.
For example, the graph Laplacian of the graph in figure \ref{fig_PolymerGraph} is given by:
\begin{equation}
L=\left(
\begin{array}{cccc}
3 & -1 & -1 & -1 \\
-1 & 3 & -1 & -1 \\
-1 & -1 & 2 & 0 \\
-1 & -1 & 0 & 2 \\
\end{array}\right).
\end{equation}

The graph Laplacian $L$ leads to the fluctuation between any pair of vertices of a Gaussian network with a given graph:
Consider the edges corresponds to Gaussian chains and the vertices have $1$-dimensional random coordinates $x=(x_1,x_2,...,x_N)$.
The elastic energy of the network is proportional to the bilinear form $x^T Lx$.
Remark that the elements of $x$ are random variables and not equal to the coordinates during randomization process.
Under the condition that the sum of the coordinates, the mass center, becomes zero: $\sum_{i=1}^N x_i=0$,
the covariance between the two elements in $x$ is given as follows \cite{jphysA2022cantarella}.
\begin{equation}
\overline{ x_a x_b} = L^+_{ab},
\end{equation}
where $+$ denotes the Moore-Penrose pseudo-inverse, which is given by the solution of the least-mean-square problem.
The fluctuation of displacement vector $x_{ab}=x_{a}-x_{b}$ between the position coordinates of vertices $a$ and $b$ is given by \cite{jphysA2022cantarella}:
\begin{equation}
\overline{\left(\Delta x_{ab}\right)^2}=\overline{\left(x_a-x_b\right)^2}=L^{+}_{aa}+L^{+}_{bb}-L^{+}_{ab}-L^{+}_{ba}.
\end{equation}

The resistance distance $\rho_{ab}$ is the effective resistance between vertices $a$ and $b$ as if the graph is an electrical circuit with each edge replaced by a unit resistor and unit current flows into the network at vertex $a$ and out of the network at $b$.
For example, the resistance distance between vertices $1$ and $2$ of the graph in figure \ref{fig_PolymerGraph} is given by $1/2$ and the resistance distance between $3$ and $4$ is given by $1$.
It is also given by \cite{jmathchem1993klein}: 
\begin{equation}
\rho_{ab}=L^{+}_{aa}+L^{+}_{bb}-L^{+}_{ab}-L^{+}_{ba}.
\label{eq_resistance}
\end{equation}
As a summary, the fluctuation between two vertices is equal to the resistance distance between them.
Remark this formulation requires only one restriction $\sum_{i=1}^N x_i=0$ and does not need the fixed boundaries \cite{jphysA2022cantarella}.

\subsection{Conditions of the simulation}
The number of vertices of a regular graph $N$ is $10648$ for all graphs that are generated.
The functionalities of regular graphs are $3$ or $4$.
We generated initial regular graphs by using Wolfram Mathematica 12.1.

We performed the following randomization process on the given initial regular graphs:
The rate of randomization $kT$ gradually decreased to the target level of $kT$ from the initial level $kT=\infty$. 
In the first stage, an initial regular graph of $22^3=10648$ vertices is embedded in the periodic $3$-dimensional $22\times 22\times 22$ cubic lattice. 
Then we applied edge switches and vertex switches to it with the rate of randomization at $kT=\infty$.
In the second stage, it was relaxed by edge switches and vertex switches at the rate $kT=20.0$.
In the third stage, it was annealed by edge switches and vertex switches where the $kT$ decreases from $kT=20.0$ to the target level, $kT=2.0, 1.0, 0.6, 0.4$, or $0.2$.
Finally, it was relaxed in the target $kT$.
The enough number of switches depends on the target $kT$ and the functionality.
The lower the rate of randomization, the larger the number of switches; the lower the functionality, the larger the number of switches.
For example, we take $160,000,000$ switches to randomized a $3$-regular graph to $kT=0.2$ and $40,000,000$ switches for a $4$-regular graph to $kT=2.0$.

We compute the pseudo-inverse of the graph Laplacian of the generated network and obtain resistance distances along an edge of the network by using intel LAPACK.

\section{Effects of cycles on the resistance distance between adjacent vertices}
\subsection{Mean resistance distance along an edge}
We define the average resistance distance along an edge for a randomized $f$-regular graph by
\begin{equation}
  \langle \rho \rangle = \frac{1}{\mathcal{E}} \sum_{(i,j)} \rho_{ij},
\end{equation}
where the symbol of sum over $(i,j)$ denotes the summation over all pairs of adjacent vertices and $\mathcal{E}$ the number of all edges in the graph.
Moreover, we also define the standard deviation of the resistance distance along an edge as follows:
\begin{equation}
  \mathrm{SD} (\rho) = \sqrt{\frac{1}{\mathcal{E}} \sum_{(i,j)} \rho_{ij}^2-{\langle \rho \rangle}^2}.
\end{equation}

It follows from Foster's theorem \cite{ccacaa2002klein} that the average resistance distance over all edges is equal to $2/f$.
For a simple graph, we obtain
\begin{equation}
\sum_{(i,j)} (LL^+L)_{ij}\rho_{ij} = -2{\rm tr} (LL^+). 
\end{equation}
We thus obtain
\begin{equation}
\sum_{i<j} L_{ij}\rho_{ij} =\mathcal{E}\langle \rho \rangle
\end{equation}
Thus, the average resistance distance along an edge is exactly $\langle \rho \rangle=(N-1)/\mathcal{E}\simeq2/f$.

The table \ref{tab_meanResi_f3} and \ref{tab_meanResi_f4} show the numerical result of the average and the standard deviation of the resistance distances of randomized $3$, $4$-regular graphs generated in the method illustrated in the previous section respectively:
\begin{table}[h]
\centering
\begin{tabular}{ccc} \hline
$kT$ & Average $\langle \rho \rangle$ & SD$(\rho)$ \\ \hline
$\infty$ & $0.666604$ & $0.00237621$ \\
$2.00$ & $0.666604$ & $0.0186843$ \\
$1.00$ & $0.666604$ & $0.0331849$ \\
$0.60$ & $0.666604$ & $0.0374478$ \\
$0.40$ & $0.666604$ & $0.0356508$ \\
$0.20$ & $0.666604$ & $0.0294013$ \\ \hline
\end{tabular}
\caption{The average resistance distances and the standard deviation of the resistance distance along an edge in $3$-regular graphs randomized at several $kT$.}
\label{tab_meanResi_f3}
\end{table}

\begin{table}[h]
\centering
\begin{tabular}{ccc} \hline
$kT$ & Average $\langle \rho \rangle$ & SD$(\rho)$ \\ \hline
$\infty$ & $0.499953$ & $0.00126746$ \\
$2.00$ & $0.499953$ & $0.0134501$ \\
$1.00$ & $0.499953$ & $0.0200116$ \\
$0.60$ & $0.499953$ & $0.0228605$ \\
$0.40$ & $0.499953$ & $0.0215961$ \\
$0.20$ & $0.499953$ & $0.0148106$ \\ \hline
\end{tabular}
\caption{The average resistance distances and the standard deviation of the resistance distance along an edge in $4$-regular graphs randomized at several $kT$.}
\label{tab_meanResi_f4}
\end{table}

We find that the resistance distance along an edge averaged over all edges is almost equal to $2/f$ for networks with $f=3,4$ and randomized at any $kT$.
The results of the average resistance distances in the tables \ref{tab_meanResi_f3},\ref{tab_meanResi_f4} are consistent with Foster's theorem.

The average resistance distance does not depend on the detailed architectures of a network.
We remark that the classical\cite{jchemphys1943james}\cite{jchemphys1947james} and improved studies\cite{science2016zhong}\cite{macro2018lang} have not recognized this behavior of the resistance distances since they derived the resistance distance along an edge in a network and they assumed that it was surrounded by infinite tree networks. It means that there are no variance for resistance distances along edges \cite{macro1989kloczkowski}, \cite{macro2018lang}, \cite{macro2019Lin}.

The result shows that the standard deviation of resistance distances depends on the randomization rate $kT$.
The resistance distance along an edge depends on the local structure surrounding the edge.
We show in the next subsection that the resistance distance along an edge is determined by the number of cycles close to the edge.
Since graphs with a low rate of randomization $kT$ have small cycles randomly, their variances of resistance distances become large.

\subsection{Effect of multiple cycles: linear dependence of the resistance distance along an edge on the number of multiple cycles around it}

We now focus on the fact that the resistance distance along an edge depends on the number of cycles around the edge. 
Since the graphs are regular and simple, the local structure around an edge is given by the sizes, numbers and locations of cycles around the edge:
we count the numbers of $3-$,$4-$ and $5$-cycles, which are separated by graph distance $d=0,1,2$ from the edge, respectively.

\begin{figure}[htbp]
\begin{center}
\includegraphics[width=10cm]{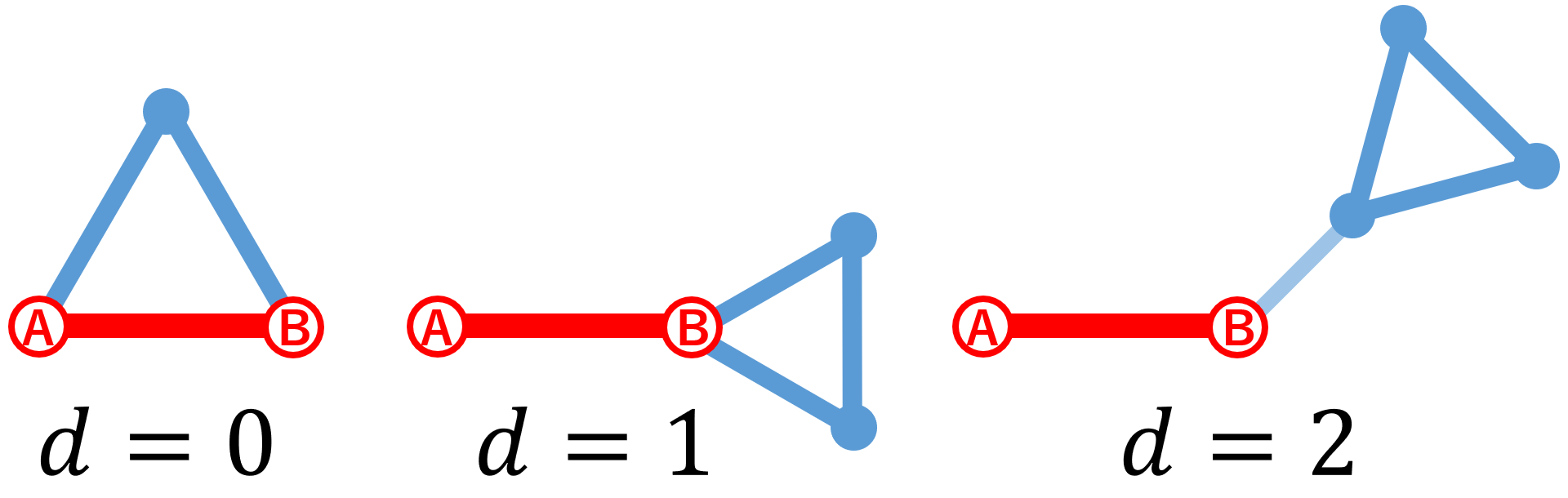}
\caption{The left panel shows the edge between a pair of vertices $A$ and $B$ has a $3$-cycle in distance $d=0$ to it. The center panel shows the edge has a $3$-cycle in distance $d=1$. The right panel shows the edge has a $3$-cycle in distance $d=2$.}
\label{fig_CycleDistance}
\end{center}
\end{figure}

We denote graph distance by $d$ as shown in fig. \ref{fig_CycleDistance}.
The existence of a cycle in graph distance $d=0$ means that the edge is included in the cycle, that of a cycle in $d=1$ means a vertex of the edge is included in the cycle, and that of a cycle in $d=2$ means a vertex of the edge is connected to the cycle via an edge.

We propose a simple linear model for the dependence of the resistance distance along an edge on the numbers of cycles around the edge:
the resistance distance along the edge is the linear sum of $2/f$ and small correction terms, we call them the effects of a single cycle.
We count the number of $3,4,5$-cycles in distance $d=0,1,2$ to an edge.
Each correction value is given by the product of the number of $k$-cycles in distance $d$ and the amplitude of the effect of a single $k$-cycle in $d$, $A_i$:
\begin{eqnarray}
\rho&=&A_0 \nonumber \\
&+&A_1\times(\#3\mathrm{-cycles\quad in\quad }d=0)
+A_2\times(\#3\mathrm{-cycles\quad in\quad }d=1) 
+A_3\times(\#3\mathrm{-cycles\quad in\quad }d=2) \nonumber \\
&+&A_4\times(\#4\mathrm{-cycles\quad in\quad }d=0) 
+A_5\times(\#4\mathrm{-cycles\quad in\quad }d=1) \nonumber \\
&+&A_6\times(\#5\mathrm{-cycles\quad in\quad }d=0).
\label{eq_linform}
\end{eqnarray}
Here, $\#3\mathrm{-cycles}$ denotes the number of $3$-cycles, $A_0$ is given by $2/f$ and $A_1, A_2,...,A_6$ are the amplitudes of the effects of a single cycle: 
the effect of a single $k$-cycle in distance $d$ is exactly derived under assumptions that an infinite regular graph should have only one $k$-cycle and the edge and the cycle should be separated by the graph distance $d$. We will show it exactly in the next subsection.

We apply the LinearModelFit of Mathematica 12.1 to the simulated values of $\rho$ and show the best estimated values for $A_0,A_1,...,A_6$ in table \ref{tab_LMFresult_f3} and \ref{tab_LMFresult_f4}.

\begin{table*}[htbp]
\begin{tabular}{cccccccc}
$kT$ & $A_0$ & $A_1$ & $A_2$ & $A_3$ & $A_4$ & $A_5$ & $A_6$\\ \hline
$\infty$ & $0.666649$ & $-0.0953498$ & $0.0472488$ & $0.0119651$ & $-0.0444645$ & $0.0220914$ & $-0.0214996$  \\
$1.00$ & $0.672475$ & $-0.0975822$ & $0.0469442$ & $0.0121814$ & $-0.0465297$ & $0.0229604$ & $-0.0241458$  \\
$0.60$ & $0.674285$ & $-0.0982084$ & $0.0468189$ & $0.0121504$ & $-0.0471503$ & $0.023429$ & $-0.0244938$  \\
$0.20$ & $0.672656$ & $-0.0954095$ & $0.0489539$ & $0.0141975$ & $-0.0479873$ & $0.0217507$ & $-0.0228524$  \\
\end{tabular}
\caption{Best estimates of parameters of the linear model for the dependence of the mean resistance distance between an edge in $3$-regular graph on the number of cycles.}
\label{tab_LMFresult_f3}
\end{table*}

\begin{table*}[htbp]
\begin{tabular}{cccccccc}
$kT$ & $A_0$ & $A_1$ & $A_2$ & $A_3$ & $A_4$ & $A_5$ & $A_6$\\ \hline
$\infty$ & $0.499984$ & $-0.0384559$ & $0.0128218$ & $0.00140745$ & $-0.0125122$ \
& $0.0041374$ & $-0.00410508$  \\
$1.00$ & $0.504491$ & $-0.0389022$ & $0.013179$ & $0.00165105$ & $-0.013324$ & $0.00480278$ & $-0.00530274$  \\
$0.60$ & $0.505894$ & $-0.0390793$ & $0.0132522$ & $0.0016829$ & $-0.013518$ & $0.00488624$ & $-0.00545694$  \\
$0.20$ & $0.505472$ & $-0.0381039$ & $0.0137314$ & $0.00192182$ & $-0.0149569$ & $0.00420389$ & $-0.00459352$  \\
\end{tabular}
\caption{Best estimates of parameters of the linear model for the dependence of the mean resistance distance between an edge in $4$-regular graph on the number of cycles.}
\label{tab_LMFresult_f4}
\end{table*}

The best estimates of $A_0,A_1,..., A_6$ well agree with the exact values of the effects of a single cycle.
For example, the exact values of the effects of a single $3$-cycle in $d=0,1,2$ in $3$-regular graphs are given by $-2/21=-0.09523, 1/21=0.04762, 1/84=0.0119$, which are the differences between $2/f$ and  the resistance distances along an edge under the assumption of a single cycle shown in the next section.

\begin{figure}[htbp]
\begin{center}
\includegraphics[width=10cm]{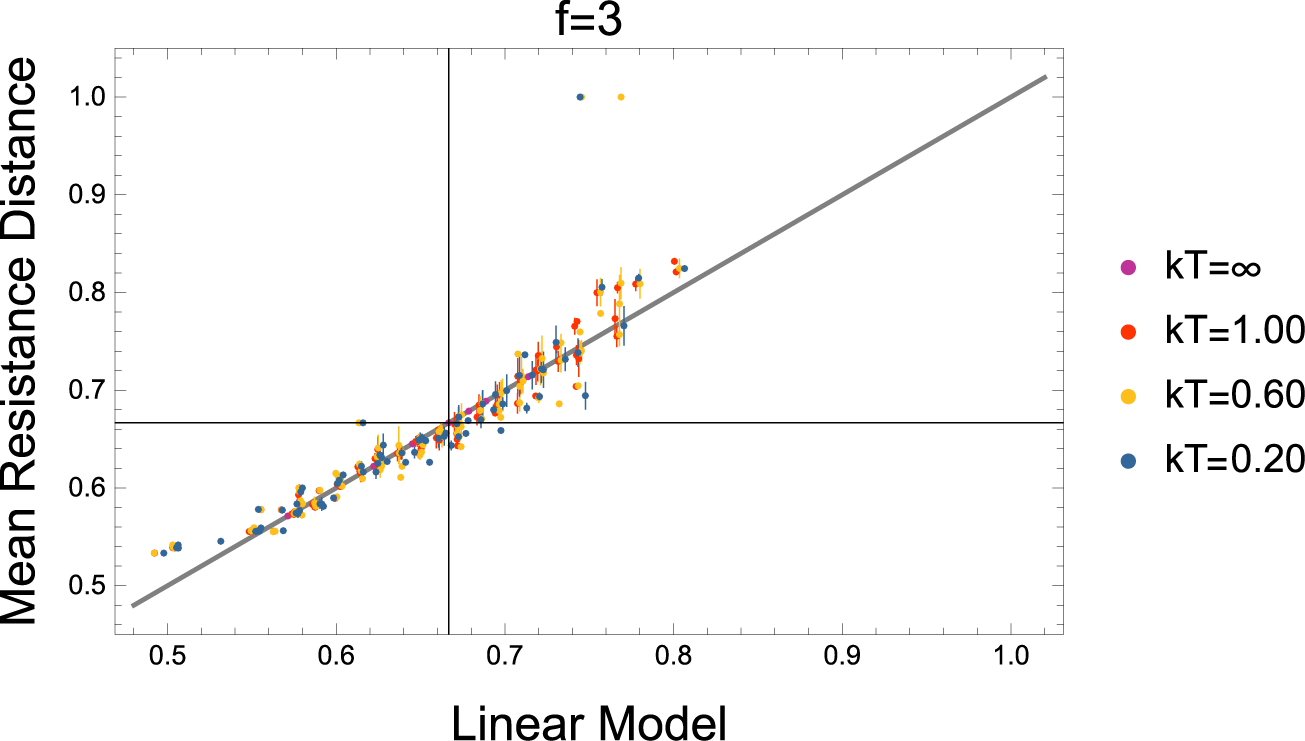}
\caption{The estimated values of the mean resistance distances of $3$-regular graphs derived with a linear model and its simulated values. The errorbars are given by their standard deviations.}
\label{fig_LMFCycleRegistance_f3}
\end{center}
\end{figure}

In figure \ref{fig_LMFCycleRegistance_f3}, we compare the estimated resistance distance with the linear model and the simulated conditional mean resistance distances in $3$-regular graphs randomized at $kT=\infty, 1.00, 0.60, 0.20$.
The horizontal axis shows the value of the linear model and the vertical axis shows the value of the conditional means calculated from the simulated results.
The solid line is $y=x$.
The error bars are given by the standard deviations of the resistance distances. 

\begin{figure}[htbp]
\begin{center}
\includegraphics[width=10cm]{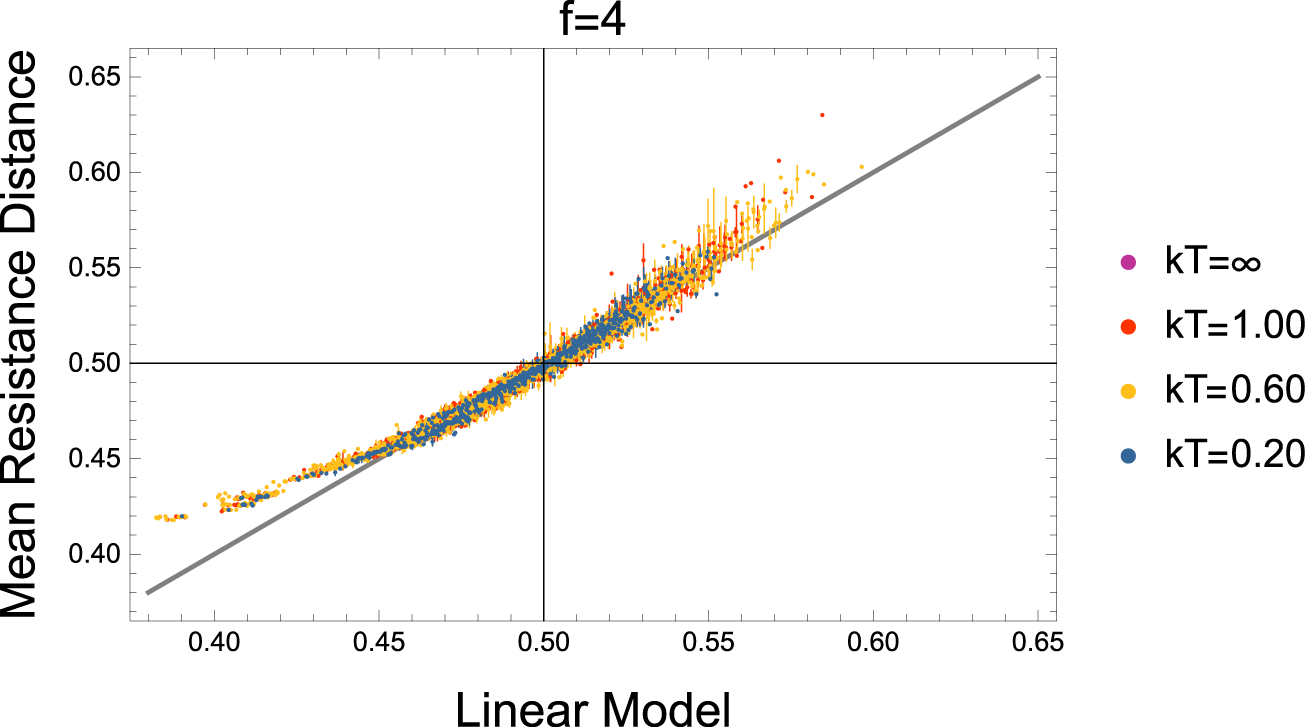}
\caption{The estimated values of the mean resistance distances of $4$-regular graphs derived with a linear model and its simulated values. The errorbars are given by their standard deviations.}
\label{fig_LMFCycleRegistance_f4}
\end{center}
\end{figure}

In figure \ref{fig_LMFCycleRegistance_f4}, we compare the estimated resistance distance with the linear model and the simulated conditional mean resistance distances of $4$-regular graphs randomized at $kT=\infty, 1.00, 0.60, 0.20$.
The horizontal axis shows the value of the same linear model and the vertical axis shows the value of the conditional means calculated from the simulated results.
The solid line is $y=x$.
The error bars are given by the standard deviations of the resistance distances. 

We discover that the conditional mean resistance distance is well expressed by the linear model in terms of the number of $k$-cycles in $d$
though figs \ref{fig_LMFCycleRegistance_f3} and \ref{fig_LMFCycleRegistance_f4} includes the conditional mean resistance distances along an edge having multiple cycles around it.
Moreover, the length of the error bars given by the standard deviation of the conditional resistance distances are quite small.
It indicates the resistance distance along an edge is determined by the number of cycles around an edge and does not depend on the other features of a network.

The agreement of the linear model with the resistance distances means that the effect of multiple cycles on the resistance distances is the linear sum of the effects of a single cycle:
Here we determine the effect of a cycle as a difference between $2/f$ and the resistance distance along an edge having the cycle around it detailed in the next subsection.
We give the exact derivation about the resistance distance along an edge having a cycle around it in the next subsection.

\subsection{Exact values of the effect of a cycle on resistance distance along an edge}
In this subsection, we show the exact values of the resistance distance along an edge having a $k$-cycle in distance $d$.
See details of the derivation in appendix.
We will find that the difference between $2/f$ and the exact resistance distance having a single cycle is very close to $A_1,A_2,...,A_6$.

For an edge included in a cycle of $k=2m+1$ vertices and there are no other cycles around it,
i.e., the edge has an odd-$k$ cycle in $d=0$,
the resistance distance along the edge is exactly given by
\begin{equation}
    \rho_{2m+1} 
    =\frac{4}{k}\sum_{b=1}^{m}\frac{1-\cos(2\pi b/k)}{2+R^*-2\cos(2\pi b/k)},
    \label{eq_Resistance_includedin_OddCycle}
\end{equation}
where $R^*$ denotes the following:
\begin{equation}
R^*=(f-2)^2/(f-1).
\end{equation}
The derivation of equation (\ref{eq_Resistance_includedin_OddCycle}) will be given in Appendix.

For an edge included in a cycle of $k=2m$ vertices, 
the resistance distance along the edge is given by 
\begin{equation}
    \rho_{2m} 
    =\frac{4}{k}\left\lbrace\sum_{b=1}^{m-1}\frac{1-\cos(2\pi b/k)}{2+R^*-2\cos(2\pi b/k)}+\frac{1}{4+R^*}\right\rbrace.
    \label{eq_Resistance_includedin_EvenCycle}
\end{equation}

When we take the limit $m\to\infty$, we obtain $4/(4+R^*)/k\to0$.
Thus, in the limit $k\to\infty$, the resistance distance along an edge having a $k$-cycle in distance $d=0$ is equal for odd $k$ and even $k$.
\begin{equation}
    \lim_{m\to\infty}\rho_{2m} 
    =4\int_0^{1/2}dx \frac{1-\cos(2\pi x)}{2+R^*-2\cos(2\pi x)}
    =1-\sqrt{\frac{R^*}{4+R^*}}
    =\frac{2}{f}.
    \label{eq_Resistance_includedin_OddCycle_lim}
\end{equation}
The limiting values of eqs. \ref{eq_Resistance_includedin_EvenCycle} and \ref{eq_Resistance_includedin_OddCycle} are equal to $2/f$.
Thus we may disregard the effect of a large-$k$ cycle on the resistance distance along an edge having the cycle in $d=0$.

We show some values of equations \ref{eq_Resistance_includedin_OddCycle} and \ref{eq_Resistance_includedin_EvenCycle} in table \ref{tab_ResistanceDistance_inCycle} for small $k$.
The first column of the table shows $k$.
The second column of the table shows the general formulae of the resistance distances for functionality $f$, the third column shows the value for $f=3$ and the forth column for $f=4$, where the real numbers after commas are the simulated values for the conditional mean and standard deviations of resistance distances along an edge included in a $k$-cycle in $f$-regular graphs.
\begin{table*}[htbp]
    \centering
    \begin{tabular}{cccc} \hline \hline
                 &  general & $f=3$ & $f=4$  \\ \hline
      &&& \\
      $3$-cycle  &  $\displaystyle\frac{2}{3+R^*}$ & $\displaystyle\frac{4}{7},\;0.571299\pm0.000688$ & $\displaystyle\frac{6}{13},\;0.461538\pm0.000073$ \\
      &&& \\ \hline
      &&& \\
      $4$-cycle  &  $\displaystyle\frac{2(3+R^*)}{(2+R^*)(4+R^*)}$ & $\displaystyle\frac{28}{45},\;0.622185\pm0.000548$ & $\displaystyle\frac{39}{80},\;0.487487\pm0.000121$ \\
      &&& \\ \hline
      &&& \\
      $5$-cycle  &  $\displaystyle\frac{2(2+R^*)}{5+5R^*+R^{*2}}$ & $\displaystyle\frac{20}{31},\;0.64515\pm0.000491$ & $\displaystyle\frac{60}{121},\;0.495849\pm0.000128$ \\
      &&& \\ \hline
      &&& \\
      $6$-cycle  &  $\displaystyle\frac{2(5+5R^*+R^{*2})}{(1+R^*)(3+R^*)(4+R^*)}$ & $\displaystyle\frac{124}{189},\;0.656085\pm0.000345$ & $\displaystyle\frac{363}{728},\;0.4986\pm0.000133$ \\
      &&& \\ \hline \hline
    \end{tabular}
    \caption{The resistance distance along an edge included in a $k$-cycle of an $f$-regular graph. Here $R^*=(f-2)^2/(f-1)$. The real numbers after commas are simulated values for the average and the standard deviation of the resistance distance along an edge included in a $k$-cycle of $f$-regular graphs randomized at $kT=\infty$.}
    \label{tab_ResistanceDistance_inCycle}
\end{table*}

For an edge having a $k$-cycle in distance $d>0$, the resistance distance along the edge is given by 
\begin{equation}
    \rho_{d}=\left(\frac{1}{1}+\frac{1}{R(d-1)+1/(f-2)}\right)^{-1}.
\end{equation}
where $R(d-1)$ is the $(d-1)$-th solution of the following recurrence relation.
\begin{equation}
\begin{split}
    R(d) &=\left(\frac{(f-2)^2}{f-1}+\frac{1}{R(d-1)+1}\right)^{-1}; \\
    R(0)   &=R_k, \\
\end{split}
\end{equation}
where $R_k$ is the resistance of a $k$-cycle.
See the exact values of $R(d-1)$ and $R_k$ in Appendix.

For finite $k$, tables \ref{tab_Resistance_ind_f3} and \ref{tab_Resistance_ind_f4} show the exact resistance distance along an edge having a $k$-cycle in distance $d>0$ and its simulated value for comparison. 
The resistance distances for a $3$-regular graph are in table \ref{tab_Resistance_ind_f3}, those for $4$-regular graphs are in table \ref{tab_Resistance_ind_f4}.
\begin{table}[htbp]
    \centering
    \begin{tabular}{cccc} \hline\hline
          & general & $d=1$ & $d=2$  \\ \hline
          &&& \\
          $3$-cycle & $\displaystyle\frac{2}{3}+\frac{4^{1-d}}{21}$ &$\displaystyle\frac{5}{7},\;0.713898\pm0.002064$ & $\displaystyle\frac{19}{28},\;0.678614\pm0.001919$  \\
          &&& \\ \hline
          &&& \\
          $4$-cycle  & $\displaystyle\frac{2}{3}+\frac{4^{1-d}}{45}$ & $\displaystyle\frac{31}{45},\;0.688741\pm0.001056$ & $\displaystyle\frac{121}{180},\;0.672184\pm0.00091$  \\
          &&& \\ \hline
          &&& \\
          $5$-cycle & $\displaystyle\frac{2}{3}+\frac{4^{1-d}}{93}$  & $\displaystyle\frac{21}{31},\;0.67737\pm0.000824$ & $\displaystyle\frac{83}{124}$, No Data  \\
          &&& \\ \hline
          &&& \\
          $6$-cycle  & $\displaystyle\frac{2}{3}+\frac{4^{1-d}}{189}$ & $\displaystyle\frac{127}{189},\;0.671945\pm0.000477$ & $\displaystyle\frac{505}{756}$, No Data  \\
          &&& \\ \hline\hline
    \end{tabular}
    \caption{Resistance distance along an edge having a $k$-cycle in distance $d$ in a $3$-regular graph.
    The real numbers after commas are the average and the standard deviation of simulated resistance distances for $3$-regular graphs randomized in $kT=\infty$. }
    \label{tab_Resistance_ind_f3}
\end{table}

\begin{table}[htbp]
    \centering
    \begin{tabular}{cccc} \hline\hline
          & general & $d=1$ & $d=2$  \\ \hline
          &&& \\
          $3$-cycle & $\displaystyle\frac{1}{2}+\frac{3^{1-2d}}{26}$ &$\displaystyle\frac{20}{39},\;0.512789\pm0.000132$ & $\displaystyle\frac{176}{351},\;0.501392\pm0.000141$  \\
          &&& \\ \hline
          &&& \\
          $4$-cycle  & $\displaystyle\frac{1}{2}+\frac{3^{1-2d}}{80}$ & $\displaystyle\frac{121}{240},\;0.50413\pm0.000174$ & $\displaystyle\frac{1081}{2160},\;0.500437\pm0.000156$  \\
          &&& \\ \hline
          &&& \\
          $5$-cycle & $\displaystyle\frac{1}{2}+\frac{3^{1-2d}}{242}$  & $\displaystyle\frac{182}{363},\;0.501352\pm0.000161$ & $\displaystyle\frac{1634}{3267}$, No Data  \\
          &&& \\ \hline
          &&& \\
          $6$-cycle  & $\displaystyle\frac{1}{2}+\frac{3^{1-2d}}{728}$ & $\displaystyle\frac{1093}{2184},\;0.500431\pm0.000154$ & $\displaystyle\frac{9829}{19656}$, No Data  \\
          &&& \\ \hline\hline
    \end{tabular}
    \caption{Resistance distance along an edge having a $k$-cycle in distance $d$ in a $4$-regular graph.
    The real numbers after commas are the average and the standard deviation of simulated resistance distances for $3$-regular graphs randomized in $kT=\infty$. }
    \label{tab_Resistance_ind_f4}
\end{table}

The formulae of the exact values of resistance distances in tables \ref{tab_Resistance_ind_f3} and \ref{tab_Resistance_ind_f4} show that the effect of a cycle on the resistance distance along an edge becomes negligible for large $k$ and $d$.
As $k$ approaches $\infty$, the limiting value of $R_k$ is $1/(f-2)$.
It follows that the resistance distance along the edge is equal to $2/f$, which is the value of the resistance distance along an edge having no cycles around it.
As $d$ increases, the expected resistance distance along an edge having a $3,4,5$-cycle in distance $d$ exponentially approaches to $2/f$.
It suggests that the effect of a cycle located in a large distance vanishes.
Since the effects of a large-$k$ cycle and that of a far cycle vanish, we only consider the effects of a short cycle near an edge for calculating resistance distance along the edge.

The exact values of resistance distances agree with the linear model:
the exact differences between $2/f$ and the exact resistance distances along an edge having $k$-cycle in distance $d\ge 0$ are close to the best estimates of $A_1,A_2,...,A_6$ in table \ref{tab_LMFresult_f3} and \ref{tab_LMFresult_f4}.
The exact resistance distances are defined on an edge under the effect of a single cycle, though the simulated edges in a regular graph have multiple cycles around itself.
The agreement of the exact differences with $A_1,...,A_6$ indicates that the effect of multiple cycles is formulated by the linear sum of the effects of single cycles assumed to be independent each other.

Let us explain the reason why the resistance distance along an edge having a cycle in $d=0$ is negative while the resistance distance along an edge having a cycle in $d>0$ is positive. 
We regard a cycle as a parallel circuit locally, where current flows into a vertex of the cycle and flows out of another vertex of it. 
Then an edge included in a cycle has a smaller resistance distance than an edge not included in any cycle. 
Furthermore, a cycle may "attract" electric current.
Let us suppose that the circuit is composed of a sub-circuit with a cycle and a sub-circuit of trees, which are connected in parallel.
If an edge is part of the sub-circuit of trees, i.e. the distance to the cycle is larger than zero $d>0$, the majority of a given current flows into the cyclic sub-circuit.
Thus, the resistance distance along the edge with a cycle in $d>0$ becomes larger than $2/f$.
Here we remark that in an electric circuit the resistance distance along an edge is given by one divided by the current.

\subsection{Distribution of resistance distances along edges and its discreteness}
The distributions of the resistance distances are not smooth. 
They have several sharp peaks with small widths, i.e., the possible values of resistance distances are discrete.

\begin{figure}[htb]
\begin{center}
\includegraphics[width=7.5cm]{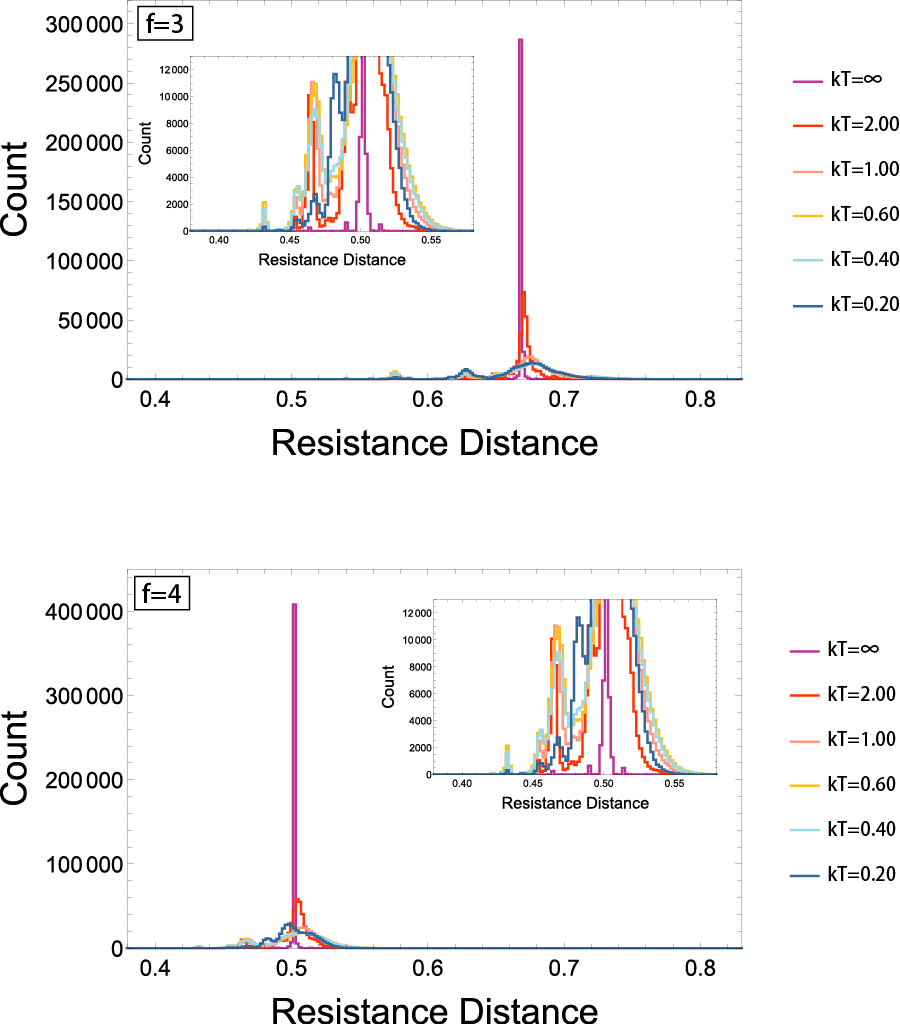}
\caption{The histograms of resistance distances between an edge of regular graphs randomized in several $kT$. The upper panel is the histograms for $3$-regular graphs and the lower panel is those for $4$-regular graphs.}
\label{fig_ResistanceHisto}
\end{center}
\end{figure}

The figure \ref{fig_ResistanceHisto} shows the histograms with bins of width $0.002$ of the resistance distances along an edge of the regular graphs randomized in the several $kT$.
The upper plot is the histogram for $3$-regular graphs and the lower one is for $4$-regular graphs.
The purple lines represent the number of edges having a resistance distance $\rho$ in the uniform regular graphs for both panels.
The red lines represent those in the regular graphs randomized in $kT=2.0$, the pink ones those in $kT=1.0$, the yellow ones those in $kT=0.6$, the pale blue ones those in $kT=0.4$ and the blue ones those in $kT=0.2$.

The highest peaks are located around at $\rho=2/f$ for all rates of randomization $kT$ and all values of functionality $f$.
The peaks of uniform $3,4$-regular graphs are prominently high and narrow.
As $kT$ becomes smaller, the peak levels become lower. 
The inner panels in figure \ref{fig_ResistanceHisto} show the enlarged view of the histograms, which show the foots of the highest peaks.
The widths of the highest peaks of $3,4$-regular graphs randomized in $kT=\infty$ are about $0.01$.
The range of peaks are very short and we can find out the tiny peaks around the highest peak.

The second highest peaks are at about $\rho=0.623$ and the third highest peaks are at $\rho=0.574$ for $3$-regular graphs randomized in low $kT$.
In contrast, for $3$-regular graphs randomized in intermediate $kT$, the second highest peaks are at $\rho=0.574$ and the third highest peaks are at $\rho=0.623$.
For uniform regular graphs, these peaks are quite low, that is, the values of the resistance distances of uniform regular graphs are concentrated to $2/f$.

The values of resistance distances are discrete.
It suggests that the resistance distance along an edge should depend on some discrete parameters of the edge: 
Since the degrees of vertices and the weights of edges are identical and the centralities of the vertices are not discrete, they shouldn't affect the resistance distances of the generated graphs.
The number of cycles around an edge is discrete and should effect the resistance distance along the edge.
We will show that in the next subsection numerically and in the next after next subsection theoretically.

\section{Frequency of short cycles}

\subsection{Dependence of the number of cycles on the rate of randomization}
\begin{figure}[ht]
\begin{center}
\includegraphics[width=12cm]{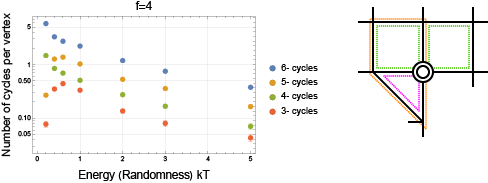}
\caption{The left panel: The mean number of cycles per vertex against the rate of randomness. The right panel: illustration on counting the number of cycles.}
\label{fig_cyclesrandomness}
\end{center}
\end{figure}

The number of cycles depends on the rate of randomization 
though the cycle rank of the network is conserved during the randomization process.
In figure \ref{fig_cyclesrandomness}, we show the mean number of cycles containing a vertex for 4-regular graphs randomized over the range from $kT=0.2$ to $5$ $kT$ in the left panel and the way how to count the number of cycles containing a vertex in the right panel.
In the left panel, the blue data points correspond to the mean numbers of $6$-cycles, the yellow ones those of $5$-cycles, the green ones those of $4$-cycles, the red ones those of $3$-cycles.
In the right panel, the vertex expressed with a double circle is contained in several cycles: It is part of a $3$-cycle (highlighted by solid magenta lines), a pair of $4$-cycles (dotted green lines), a $5$-cycle (orange dashed line). 
The number of $3$-cycles containing the vertex is $1$, that of $4$-cycles is $2$ and that of $5$-cycles is $1$.

The mean number of cycles containing a vertex decreases as $kT$ increases.
Remark for the low rate of randomization, $kT=0.2$, the mean numbers of $3,5$-cycles approach zero while those of $4,6$-cycles become larger monotonically.
The reason is that the graphs were embedded in a cubic lattice during randomization.
Randomization at low $kT$ replaces diagonal edges between next-neighboring vertices by vertical or horizontal edges between neighboring vertices on lattice.
Thus the mean number of odd cycles approaches $0$ as $kT\to0$.

The standard deviations of the resistance distances listed in tables \ref{tab_meanResi_f3} and \ref{tab_meanResi_f4} increase as $kT$ decreases,
since the number of cycles depends on $kT$ and it decreases as $kT$ decreases.
If $kT=\infty$, a network has few short cycles. Thus, an edge will be surrounded by $(f-1)$-ary trees.  
As shown in the previous section, the effect of cycles decreases exponentially as $d$ grows. Thus, the effect of cycles on resistance distance for most edges in the network is equal to zero and most edges have resistance distance $2/f$.
This is the reason why the standard deviations of the resistance distance are quite small. 
If $kT$ is small, a network has many short cycles. 
It makes the standard deviations of resistance distance larger.

\subsection{Probabilities of links between vertices neighboring on the template coordinates}
The mean number of short cycles containing an edge decreases as $kT$ increases and becomes almost zero when $kT=\infty$.
If $kT=\infty$, the probability of a pair of vertices having an edge between them is $f/N$ for any pair of vertices in the network.
Thus, if vertices $A$ and $B$ are connected and $B$ and $C$ are connected, the probability of the network having an $(A-B-C-A)$ cycle is $(f-1)/N$.
Therefore, when $N$ is large, there are few $3$-cycles in a network randomized at $kT=\infty$.

After randomization at low $kT$, the geometrical close vertices will be connected with much larger probability than $f/N$.
If vertices $A$ and $B$ are connected and vertices $B$ and $C$ are connected, it means $A$ and $B$, $B$ and $C$ are geometrically close.
Thus also $A$ and $C$ are geometrically close.
$A$ and $C$ have an edge with larger probability than $(f-1)/N$.
As a result, a randomized network at a low $kT$ has more $3$-cycles than a network randomized at high $kT$.
This suggests that the statistical properties of networks are different for the different rate of randomization $kT$ though the simple networks have the same number of vertices $N$ and the same functionality $f$.

Table \ref{tab_AdjacentLinks} shows the probabilities of edges connecting between a pair of vertices neighboring on the template coordinates during randomization.
Networks randomized at $kT=\infty$ have few edges connecting neighboring vertices.
In contrast, edges in networks randomized at low $kT$ connect a pair of neighboring vertices with larger probability than that for networks in high $kT$.
\begin{table}[htbp]
    \centering
    \begin{tabular}{ccccccc} \hline\hline
         & $kT=\infty$ & $kT=2.0$ & $kT=1.0$ & $kT=0.6$ & $kT=0.4$ & $kT=0.2$ \\ \hline
         $f=3$ & $0.06\%$ & $24.33\%$ & $46.42\%$ & $67.30\%$ & $83.44\%$ & $97.19\%$ \\ \hline
         $f=4$ & $0.04\%$ & $23.44\%$ & $44.88\%$ & $64.51\%$ & $80.89\%$ & $97.78\%$ \\ \hline\hline
    \end{tabular}
    \caption{Ratio of edges connecting between neighboring vertices on the template coordinates during randomization.}
    \label{tab_AdjacentLinks}
\end{table}
These probabilities of links between neighboring vertices $p_l$ is approximately given by $1/p_l = 0.95 + 1.13(kT)^{1.5}$ for $f=3$ and $1/p_l=0.97 + 1.18(kT)^{1.5}$ for $f=4$ from the data of Table \ref{tab_AdjacentLinks}. 
The dependence of $1.5$-th power of $kT$ should be explained as follows. The mean square length of an edge during randomization $\langle (x(V)-x(W))^2 \rangle$ on the template coordinates is proportional to $kT$.
A vertex tends to be connected to vertices in a sphere of radius $\sqrt{\langle (x(V)-x(W))^2 \rangle}$ centered at the coordinate of the vertex.
Thus $p_l$ is proportional to $1/(\sqrt{\langle (x(V)-x(W))^2 \rangle})^3 =1/kT^{1.5} $.

\section{Conclusions} 

We have derived the resistance distance along an edge having a $k$-cycle in distance $d$ in an $f$-regular graph network and have numerically shown that the resistance distance along an edge having multiple cycles around it is expressed as the linear formula (\ref{eq_linform}) of the number of cycles around the edge and the functionality $f$.
The electrical circuit analogy enables us to derive the asymptotic formula of the resistance distance along an edge having a cycle around it in the infinite large network.
It elucidates how a small cycle changes resistance distances along edges around the cycle:
First, an edge included in the cycle has a smaller resistance distance than $2/f$ because the cycle forms a parallel circuit locally.
Second, an edge next to the cycle has a larger resistance distance than $2/f$ because the cycle "attracts" the electric current and the edge has smaller current.

In addition, we suggest that the behavior of a small cycle that takes current away from the surrounding edges generally increases a resistance distance between a pair of vertices separated by several edges, i.e., their graph distance is larger than $1$.
For instance, a pair of vertices separated by $2$ edges cannot be included in a $3$-cycle. 
Thus a $3$-cycle should only enlarge the resistance distance between the pair of vertices.
We expect that in a network with a number of small cycles, the resistance distance between a pair of vertices separated by edges becomes large.
It may change the statistical properties of the polymer network.

We have also shown that the number of cycles around an edge depends on the rate of randomization.
Homogeneous networks can have different densities of cycles even if their numbers of vertices are identical and their vertices have the same degrees of valency $f$ in common.
We conclude that if we modify the rate of randomization, the density of cycles changes and several statistical properties of the homogeneous networks are affected.


\section{Acknowledgements}

We would like to thank Jason Cantarella and Clayton Shonkwiler for introducing Foster's theorem to our research and for detailed discussion on the resistance distance of a regular graph.
We are also grateful to the Japan Science and Technology Agency (CREST Grant No. JPMJCR19T4).

\appendix
\section{Appendix}

In this appendix, we show that the resistance distance along an edge  depends on the existence of a short cycle around it.

\subsection{Resistance distance along an edge surrounded by trees}

\begin{figure}[htbp]
\begin{center}
\includegraphics[width=10cm]{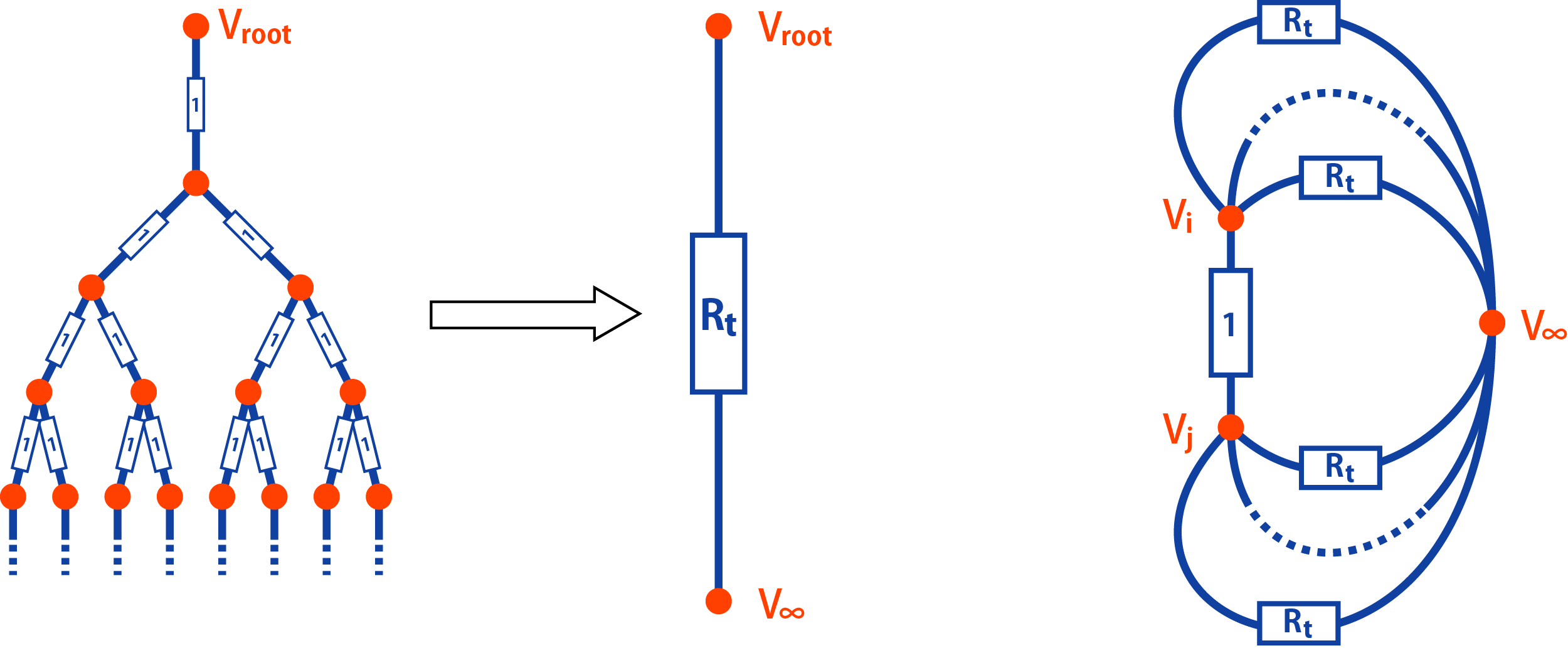}
\caption{The left diagrams show that a tree circuit is reduced to a simple circuit with a resistance $R_{t}$. Red disks express nodes and white rectangles express resistances.
The right diagram is a reduced circuit around the vertices $V_i$ and $V_j$ connected with an edge of unit resistance in a regular graph. In the left diagram, $V_i$ and $V_j$ are incident to the virtual vertex $V_{\infty}$ via $f-1$ edges of resistance $R_t$ respectively.}
\label{fig_Fary1Tree}
\end{center}
\end{figure}

Consider the resistance between the root and the ends of a tree graph consisting of a complete $(f-1)$-ary tree and an edge incident to the root of the tree as shown in the left diagram of figure \ref{fig_Fary1Tree}.
Each edge has a unit resistance.
Current flows into the root of the tree and out of the ends of the tree.
The resistance along the tree with $d$ levels depth is given by $\sum_{i=0}^{d}(f-1)^{-i}$.
The sum exponentially approaches to the asymptotic value given by
\begin{equation}
    R_{t}=\sum_{i=0}^{\infty}\left(\frac{1}{f-1}\right)^i=\frac{f-1}{f-2}.
\end{equation}
We reduce the tree into an edge having resistance $R_{t}$.
The edge is from the root vertex of the tree $V_{root}$ to a virtual vertex $V_{\infty}$ replacing all ends of the tree.

We now estimate the resistance distance between vertices $V_i$ and $V_j$ connected with an edge in an $f$-regular graph. 
We assume that the regular graph has few cycles and thus $V_i$ and $V_j$ are surrounded by trees.
Vertices $V_i$ and $V_j$ are incident to $f-1$ trees, respectively. 
Now we replace each tree by an edge of resistance $R_{t}$. Their ends are incident to the same virtual vertex $V_{\infty}$ shown in the right diagram of the figure \ref{fig_Fary1Tree}.
Moreover, we reduce the circuit again: $f-1$ trees are replaced by an edge of resistance $R_1 \equiv R_{t}/(f-1)$.
As a result, the resistance distance between $V_i$ and $V_j$ is given by
\begin{equation}
    \rho_{ij}=\left(\frac{1}{1}+\frac{1}{2R_{1}}\right)^{-1}   
    =\frac{2}{f}.
\end{equation}
Therefore, the resistance distance between a pair of vertices connected by an edge in a regular graph is almost equal to $2/f$ if they have few cycles around themselves.
This value is identical to the average resistance distance derived via Foster's theorem.

Here we remark that Kloczkowski et al.\cite{macro1989kloczkowski} derived a similar result for the edges near the center of a tree graph where all boundary vertices are fixed to the ground.

\subsection{Resistance distance along an edge included in a $k$-cycle}

\begin{figure}[htbp]
\begin{center}
\includegraphics[width=6cm]{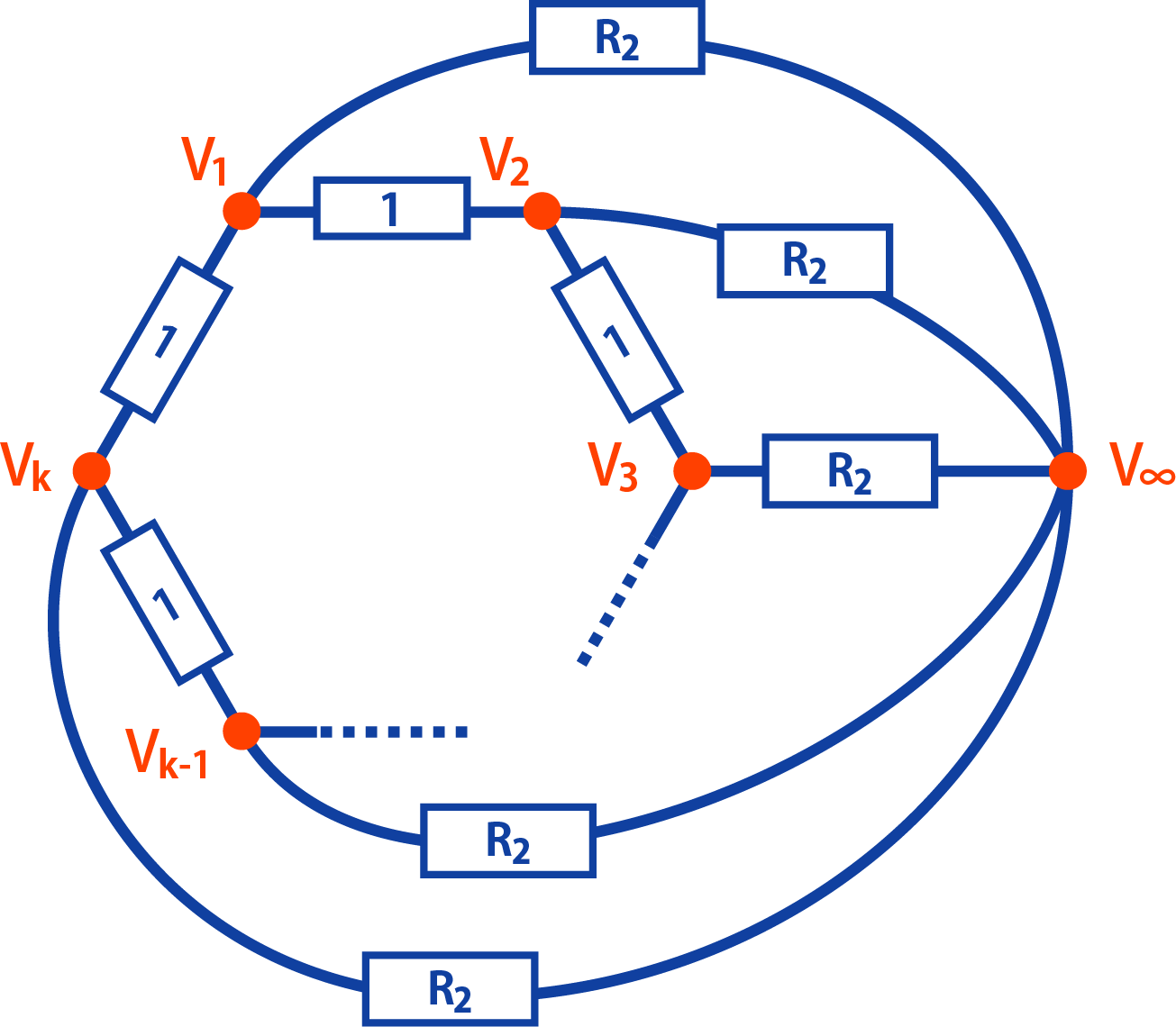}
\caption{Reduced diagram of a $k$-cycle in an $f$-regular graph. The cycle consists of $k$ vertices labeled $V_1,V_2,V_3,...,V_{k-1},V_k$ and $k$ edges of unit resistance. The vertices are incident to the same virtual vertex $V_{\infty}$ via an edge of resistance $R_2=R_t/(f-2)$.}
\label{fig_ResistanceinCycle}
\end{center}
\end{figure}

Consider the resistance distance along an edge included in a $k$-cycle consisting of vertices $V_1,V_2,V_3,...,V_{k-1},V_k$ in an $f$-regular graph as shown in figure \ref{fig_ResistanceinCycle}.
Each pair of the adjacent vertices of the cycle is connected by an edge of unit resistance.
Each vertex of the cycle is incident to the same virtual vertex $V_{\infty}$ via an edge replacing $(f-1)$-ary trees of resistance $R_2\equiv R_t/(f-2)$ .

We compute the resistance distance of the circuit by using a weighted graph Laplacian $L'$.
The $(i,j)$ off-diagonal element of the weighted graph Laplacian for a electrical circuit $L'_{ij}$ is given by $-1/R_{ij}$ where the $i$-th and $j$-th nodes are connected by an edge of resistance $R_{ij}$.
The $i$-th diagonal element $L'_{ii}$ is given by $-\sum_{i\neq j}L'_{ij}=\sum_{i\neq j}1/R_{ij}$.
The resistance distance between the $i$-th and $j$-th vertices are given by
\begin{equation}
\rho_{ij}=L'^+_{ii}-L'^+_{ij}-L'^+_{ji}+L'^+_{jj}
\end{equation}

For the circuit $V_1,V_2,...,V_k$ and $V_{\infty}$ shown in figure \ref{fig_ResistanceinCycle}, the weighted graph Laplacian is given by
\begin{equation}
\left(
\begin{array}{cccccccc}
 2+R^* & -1 & 0 & \cdots & \cdots & 0 & -1 & -R^* \\
-1 & 2+R^* & -1 & 0  & \cdots & \cdots & 0 & -R^*  \\
 0 & -1 & 2+R^* & -1 & & \cdots & 0 & -R^* \\
 \vdots  & 0  & -1 & 2+R^* & & & \vdots & \vdots \\
 \vdots  & \vdots & & & \ddots & & 0 & -R^*\\
 0 & \vdots & \vdots & & & & -1 & -R^*\\
-1 & 0 & 0 & \cdots & 0 & -1 & 2+R^* & -R^* \\
-R^* & -R^* & -R^* & \cdots & -R^* & -R^* & -R^* & kR^*
\end{array}\right),
\end{equation}
here we abbreviate $1/R_2$ to $R^*$.

The weighted graph Laplacian $L'$ for the graph has $k+1$ eigenvectors:
One of them is $(1,1,...,1)/\sqrt{k+1}$ with eigenvalue $0$. 
Another eigenvector is $(-1,-1,...,-1,k)/\sqrt{k(k+1)}$ with eigenvalue $(k+1)R^*$.
The other eigenvectors have complex exponentials in the $1$st through the $k$-th elements and $0$ in the last element: $(\exp(2\pi in/k),\exp(4\pi in/k),\exp(6\pi in/k),...,\exp(2\pi ink/k),0)/\sqrt{k}$ for $n=\pm 1,\pm 2, ...$
Their eigenvalues are given by $2+R^*-2\cos(2\pi n/k)$.

The $(i,j)$ element of the pseudo-inverse of $L'$ is given by
\begin{equation} \label{eqlampseudo}
    L'^+_{ij}=\left(U'\Lambda'^+ U^\dagger\right)_{ij}=\sum_{a=1}^{k}\frac{1}{\lambda'_a}U'_{ia}\bar{U}'_{ja},
\end{equation}
where the first column of $U'$ is $(-1,-1,...,-1,k)/\sqrt{k(k+1)}$,
the last column of $U'$ is $(1,1,...,1)/\sqrt{k+1}$,
and the other columns are $(\exp(2\pi in/k),\exp(4\pi in/k),\exp(6\pi in/k),...,\exp(2\pi ink/k),0)/\sqrt{k}$ for $n=\pm 1, \pm 2,...$.
In eq.(\ref{eqlampseudo}), $\Lambda^{+}$ is the diagonal matrix whose $i$-th diagonal is the inverse of the eigenvalue corresponding to the $i$-th eigenvector or zero if the eigenvalue is zero.

In the rest of this subsection, we compute the resistance distance between vertices $V_1$ and $V_2$, i.e., consider the case for $(i,j)=(1,2)$.
Remark that resistance distance is identical for every pair of adjacent vertices in the cycle, $(V_1,V_2),(V_2,V_3),...,(V_k,V_1)$, because of the symmetry of the circuit.

For an edge included in an odd $k=2m+1$ cycle, the elements of the pseudo-inverse of the weighted reduced graph Laplacian are given by 
\begin{equation}
    L'^+_{11}=\frac{1}{k(k+1)^2 R^*}+\frac{1}{k}\sum_{b=1}^{m}\frac{2}{2+R^*-2\cos(2\pi b/k)}
\end{equation}
and
\begin{equation}
    L'^+_{12}
    =\frac{1}{k(k+1)^2 R^*}+\frac{1}{k}\sum_{b=1}^{m}\frac{2\cos(2\pi b/k)}{2+R^*-2\cos(2\pi b/k)}.
\end{equation}
The resistance distance along the edge is given by
\begin{equation}
    \rho_{12} 
    =2L'^+_{11} -2L'^+_{12}
    =\frac{4}{k}\sum_{b=1}^{m}\frac{1-\cos(2\pi b/k)}{2+R^*-2\cos(2\pi b/k)}.
    \label{eq_ResistanceinOddCycle}
\end{equation}
Here we use $L'^+_{22}=L'^+_{11}$ and $L'^+_{21}=L'^+_{12}$.

For an edge included in an even $k=2m$ cycle, the elements of the pseudo-inverse of the weighted reduced graph Laplacian are given by 
\begin{equation}
    L'^+_{11}
    =\frac{1}{k(k+1)^2 R^*}+\frac{1}{k}\sum_{b=1}^{m-1}\frac{2}{2+R^*-2\cos(2\pi b/k)}+\frac{1}{k(4+R^*)}
\end{equation}
and
\begin{equation}
    L'^+_{12}
    =\frac{1}{k(k+1)^2 R^*}+\frac{1}{k}\sum_{b=1}^{m-1}\frac{2\cos(2\pi b/k)}{2+R^*-2\cos(2\pi b/k)}-\frac{1}{k(4+R^*)}.
\end{equation}
Remark the $k$-th eigenvector is $(\exp(2\pi im/k),\exp(4\pi im/k),...,0)/\sqrt{k}=(1,-1,1,-1,...,-1,0)/\sqrt{k}$.
The resistance distance along the edge is given by 
\begin{equation}
    \rho_{12} 
    =\frac{4}{k}\left\lbrace\sum_{b=1}^{m-1}\frac{1-\cos(2\pi b/k)}{2+R^*-2\cos(2\pi b/k)}+\frac{1}{4+R^*}\right\rbrace.
    \label{eq_ResistanceinEvenCycle}
\end{equation}

For several rates of randomization,
the average resistance distance along an edge included in a $k$-cycle in an $f$-regular graph is consistent with the values given by eqs. (\ref{eq_ResistanceinOddCycle}) and (\ref{eq_ResistanceinEvenCycle}). 
For an edge included in a $3$-cycle of a $3$-regular graph randomized in $kT=1.00$, the average simulated resistance distance is $0.573183\pm0.003048$, that for $kT=0.20$ is $0.573798\pm0.003887$.
For an edge included in a $4$-cycle of a $3$-regular graph randomized in $kT=1.00$, the average simulated resistance distance is $0.624671\pm0.0043461$, that for $kT=0.20$ is $0.62517\pm0.005695$.
Here the errors are the standard deviations of the resistance distances.
The standard deviations are quite small and it implies that the resistance distance along an edge depends on an existence of a cycle close to the edge and is not disturbed by the other properties of the graph.
The simulated values are also close to the values in the table \ref{tab_ResistanceDistance_inCycle} but slightly increase as $kT$ decreases.
As shown in the following section, the number of cycles increases as $kT$ decreases.
And as shown in the previous section, if short cycles exist around an edge but do not include the edge, the resistance distance along the edge increases.
Thus $kT$ decreases, the average resistance distance along an edge included in a short cycle slightly increases.

\subsection{Resistance distance of an edge not included in a short cycle but adjacent to a $k$-cycle}
\begin{figure}[htbp]
\begin{center}
\includegraphics[width=12cm]{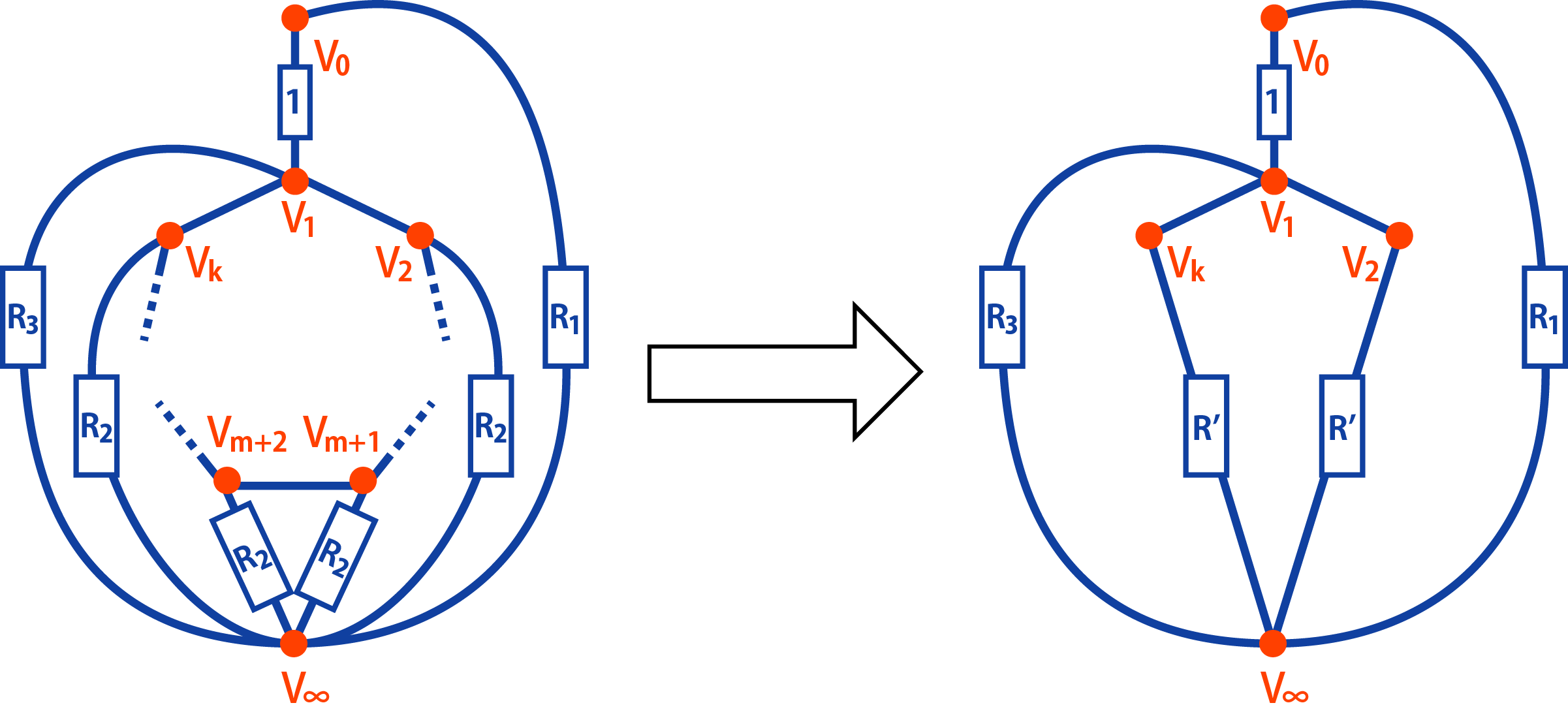}
\caption{The reduced diagrams of an edge incident to an odd-k cycle in an $f$-regular graph. 
In the left diagram, the cycle consist of $k$ edges $(V_1-V_2),(V_2-V_3),...,(V_{k-1}-V_k),(V_k-V_1)$ of unit resistance. The vertices $V_2,V_3,...,V_k$ are incident to a virtual vertex $V_{\infty}$ through the reduced edges of resistance $R_2=R_t/(f-2)$. If $k>3$, $V_1$ is incident to $V_{\infty}$ through the reduced edge of resistance $R_3=R_t/(f-3)$. The vertices $V_0$ and $V_1$ are connected with an edge of unit resistance. $V_0$ is also connected with $V_{\infty}$ through the reduced edge of resistance $R_1=R_t/(f-1)$.
The right diagram is the reduced diagram of the left one. The vertices $V_3,V_4,...,V_{k-1}$ are removed. $V_2$ and $V_k$ are incident to $V_{\infty}$ through the reduced edges of resistance $R'$ respectively.}
\label{fig_Resistance_NeighborCycle_odd}
\end{center}
\end{figure}

We compute the resistance distance along an edge next to a $k$-cycle.
As in the previous subsection, the $k$-cycle consists of $k$ vertices, $V_1,V_2,...,V_k$ and $k$ edges of unit resistance.
Vertex $V_0$ is connected to $V_1$ with an edge of unit resistance as shown in the figure \ref{fig_Resistance_NeighborCycle_odd}.
The vertex $V_2,V_3,...,V_k$ is connected to the virtual vertex $V_{\infty}$ via a reduced edge of resistance $R_2=R_t/(f-2)=(f-1)/(f-2)^2$.
If $f>3$, $V_1$ is connected to $V_{\infty}$ via a reduced edge of resistance $R_3=R_t/(f-3)=(f-1)/(f-2)(f-3)$ and $V_0$ is connected to it via a reduced edge of resistance $R_1=R_t/(f-1)=1/(f-2)$.

If the number of vertices of the cycle is odd, $k=2m+1$, and if current flows into $V_0$ out of $V_1$, the edge between $V_{m+1}$ and $V_{m+2}$ can be removed because of the symmetry of the circuit.
After the removal of the edge, we reduce the edges in the cycle recursively:
(i) We replace $V_{\infty}-V_{m+1}-V_{m}$ edges by a new edge $V_{\infty}-V_{m}$ of resistance $1+R_2$.
Then we replace the two parallel edges from $V_{m}$ to $V_{\infty}$ by an edge of resistance
\begin{equation}
    \left(\frac{1}{R_2}+\frac{1}{R_2+1}\right)^{-1}
    =\frac{R_2(1+R_2)}{2R_2+1}.
\end{equation}
(ii) After $m-i-1$ recursions, the resistance between $V_{i-1}$ and $V_{\infty}$ is $R_{i-1}$ given by
\begin{equation}
\begin{split}
    R_{i-1}
    &=\frac{R_2(1+R_i)}{R_i+R_2+1}=\frac{R_2 R_i + R_2}{R_i+R_2+1}; \\
    R_{m+1}&=R_2.
\end{split}
\label{eq_recursion_NeighborCycle}
\end{equation}
(iii) After $m-1$ recursions, the vertices $V_3,V_4,...,V_{m+1}$ are removed and $V_2$ is connected to $V_{\infty}$ through an edge of resistance $R'$. 
The vertices $V_k,V_{k-1},...,V_{m+2}$ are also removed and $V_k$ is connected to $V_{\infty}$ through an edge of resistance $R'$ because of the circuit symmetry as shown in the right diagram in figure \ref{fig_Resistance_NeighborCycle_odd}.
Resistance $R'$ is given by
\begin{equation}
    R'=\frac{\alpha_{+}Q_{m-1}-\alpha_{-}}{Q_{m-1}-1},
\end{equation}
where
\begin{equation}
\begin{split}
    \alpha_{\pm} &\equiv \frac{-1\pm\sqrt{1+4R_2}}{2},\\
    Q_{j}&=\frac{R_2-\alpha_{-}}{R_2-\alpha_{+}}Q_{j-1}=\left(\frac{R_2-\alpha_{-}}{R_2-\alpha_{+}}\right)^{j} Q_0,\\
    Q_0&=\frac{R_2-\alpha_{-}}{R_2-\alpha_{+}}.
\end{split}
\end{equation}
We obtain the explicit term of $R'$ as
\begin{equation}
    R'=\frac{1}{2}\left(\sqrt{4R_2+1}-1+\frac{2\sqrt{4R_2+1}}{\left(\frac{2R_2+1+\sqrt{4R_2+1}}{2R_2+1-\sqrt{4R_2+1}}\right)^m-1}\right).
\end{equation}
Since $\sqrt{4R_2+1}=f/(f-2)$, the term is rational.
It becomes
\begin{equation}
    R'=\frac{1}{f-2}\left(1+\frac{f}{(f-1)^{2m}-1}\right).
\end{equation}
We remark that the limiting value of the formula as $m$ approaches $\infty$ is $1/(f-2)$.

For example, when $m=1$, i.e., $k=3$, $R'=R2$, when $m=2$,$R'=R_2(R_2+1)/(2R_2+1)$, when $m=3$, $R'=R_2(R_2^2+3R_2+1)/(1+R_2)(2+3R_2)$.
The resistance distance between $V_0$ and $V_1$ is
\begin{equation}
    \rho_{01}
    =\left(\frac{1}{1}+\frac{1}{R_1+\left(2/(R'+1)+(f-3)/R_t\right)^{-1}}\right)^{-1}.
\end{equation}

When $k=3$, it is given by
\begin{equation}
    \rho_{01}=\frac{2}{f-1}-\frac{2}{1+f(f-1)}.
\end{equation}
For $f=3$, $\rho_{01}$ is $5/7$, for $f=4$, $20/39$.
They are close to the simulated values.

For the number of vertices of the cycle is even, $k=2m$, we deform the circuit to the odd-$k$ circuit by removing the vertex $V_{m+1}$ via Y-$\Delta$ transform.
The Y-$\Delta$ transformation removes $V_{m+1}$ and the $3$ edges incident to it and creates $3$ new edges between $V_{\infty}$ and $V_{m}$, $V_{\infty}$ and $V_{m+2}$, $V_{m}$ and $V_{m+2}$ of resistance $2R_2+1$, $2R_2+1$ and $(2R_2+1)/R_2$ respectively.
Then we can contract the edge chains $V_2-V_3...-V_m$ and $V_k-V_{k-1}-...-V_{m+2}$ to edges of resistance $R'$ respectively via recursive operation as in the calculation for an odd-$k$ cycle.
The contracted resistance $R'$ for an even $k$ is given by
\begin{equation}
    R'=\frac{\alpha_{+}Q'_{m-2}-\alpha_{-}}{Q'_{m-2}-1},
\end{equation}
where
\begin{equation}
\begin{split}
    Q'_{j}&=\frac{R_2-\alpha_{-}}{R_2-\alpha_{+}}Q'_{j-1}=\left(\frac{R_2-\alpha_{-}}{R_2-\alpha_{+}}\right)^{j} Q'_0,\\
    Q'_0&=\frac{\left(\frac{1}{R_2}+\frac{1}{2R_2+1}\right)^{-1}-\alpha_{-}}{\left(\frac{1}{R_2}+\frac{1}{2R_2+1}\right)^{-1}-\alpha_{+}}.
\end{split}
\end{equation}
We obtain the followings:
\begin{equation}
    R'=\frac{1}{f-2}\left(1+\frac{f}{(f-1)^{2m-1}-1}\right).
\end{equation}
We remark that the limiting value of the formula as $m$ approaches $\infty$ is also $1/(f-2)$.
It is identical to the case of odd $m$.

For example, when $k=4$, it is given by
\begin{equation}
    \rho_{01}=\frac{2}{f^2}\left(f+\frac{1}{f^3-3f^2+4f-2}\right).
\end{equation}
For $f=3$, it is $31/45$, $f=4$, $121/240$.
They are close to the simulated value.

\subsection{Resistance distance along an edge having a $k$-cycle in distance $d$}

\begin{figure}[htbp]
\begin{center}
\includegraphics[width=8cm]{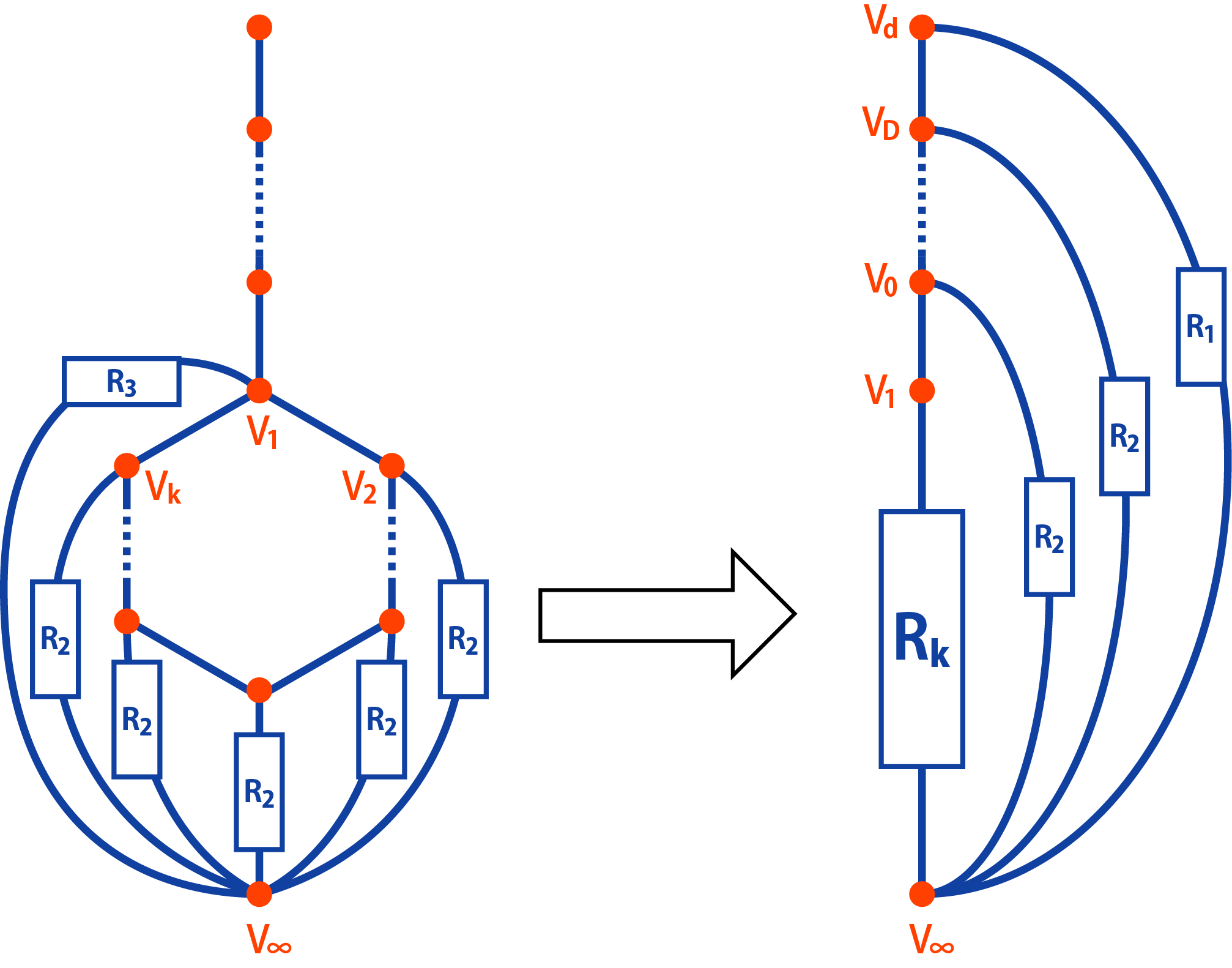}
\caption{The edge $V_d-V_D$ having a $k$-cycle in distance $d$. The edge has unit resistance. The $k$-cycle consists of $k$ vertices $V_1,V_2,...,V_k$ and $k$ edges of unit resistance.
In the left diagram, each vertex of $V_2,V_3,...,V_k$ is connected to the virtual vertex $V_{\infty}$ with an edge of resistance $R_2=R_t/(f-2)$. If $f>3$, $V_1$ is connected to $V_{\infty}$ with an edge of resistance $R_3=R_t/(f-3)$.
The right diagram is a reduced diagram of the left one. The vertices  $V_2,V_3,...,V_k$ are removed and $V_1$ is connected to $V_{\infty}$ with a contracted edge of resistance $R_k$.
$V_d$ is connected to $V_{\infty}$ with an edge of resistance $R_1=R_t/(f-1)$. The other vertices are connected to $V_{\infty}$ with  edge of resistance $R_2$ respectively.}
\label{fig_DistanceCycle}
\end{center}
\end{figure}

We calculate the resistance distance along an edge having a $k$-cycle in distance $d$:
The edge has a pair of vertices in its end, $V_D$ and $V_d$.
As in the previous subsection, the $k$-cycle consists of $k$ vertices labeled $V_1,V_2,...,V_k$ and the vertex $V_0$ is incident to $V_1$ as illustrated in the left diagram of the figure \ref{fig_DistanceCycle}.
The vertices $V_2,V_3,...,V_k$ are connected to a virtual vertex $V_\infty$ with edges of resistance $R_2=R_t/(f-2)$. 
If $f>3$, $V_1$ is connected to $V_{\infty}$ with an edge of resistance $R_3=R_t/(f-3)$.
Each pair of vertices $(V_1,V_2),(V_2,V_3),...,(V_k,V_1)$ is connected with an edge of unit resistance and the edge $(V_0-V_1)$ has unit resistance.
The graph distance between $V_d$ and $V_1$ is $d$ and the graph distance between $V_D$ and $V_0$ is $d-2$.

For calculating the resistance distance, we reduce the edges in the $k$-cycle as shown in the right diagram of the figure \ref{fig_DistanceCycle}:
The vertices  $V_2,V_3,...,V_k$ are removed and $V_1$ is connected to $V_{\infty}$ with a reduced edge of resistance $R_k$ given by
\begin{equation}
    R_k=\left(\frac{f-3}{R_t}+\frac{2}{R'+1}\right)^{-1}
\end{equation}
We remark the limiting value of $R_k$ as $k$ approaches $\infty$ is $1/(f-2)=R_1$.

$V_d$ is connected to $V_{\infty}$ with an edge of resistance $R_1=R_t/(f-1)$. 
The other vertices are connected to $V_{\infty}$ with edges of resistance $R_2$ respectively.

We calculate the resistance distance between $V_d$ and $V_D$ via recursive operation.
Let $R_l$ be the resistance between $V_{\infty}$ and the vertex in distance $l$ from $V_1$:
\begin{equation}
\begin{split}
    R_{l+1} &=\left(\frac{1}{R_2}+\frac{1}{R_l+1}\right)^{-1}=\frac{R_2 R_l+R_2}{R_l+R_2+1} \\
    R_0   &=R_k \\
\end{split}
\end{equation}
This equation is almost same to the equation \ref{eq_recursion_NeighborCycle}.
Thus we obtain
\begin{equation}
    R_{l}
=\frac{\alpha_{+}Q'_{l}-\alpha_{-}}{Q'_{l}-1},
\end{equation}
where
\begin{equation}
\begin{split}
    Q'_{l}&=\left(\frac{R_2-\alpha_{-}}{R_2-\alpha_{+}}\right)^{l} Q'_0,\\
    Q'_0&=\frac{R_k-\alpha_{-}}{R_k-\alpha_{+}}.
\end{split}
\end{equation}

The resistance distance between $V_d$ and $V_D$ is
\begin{equation}
    \rho_{dD}=\left(\frac{1}{1}+\frac{1}{R_{d-1}+R_1}\right)^{-1}.
\end{equation}

\subsection{Mean Graph Distances}

\begin{figure}[ht]
\begin{center}
\includegraphics[width=8cm]{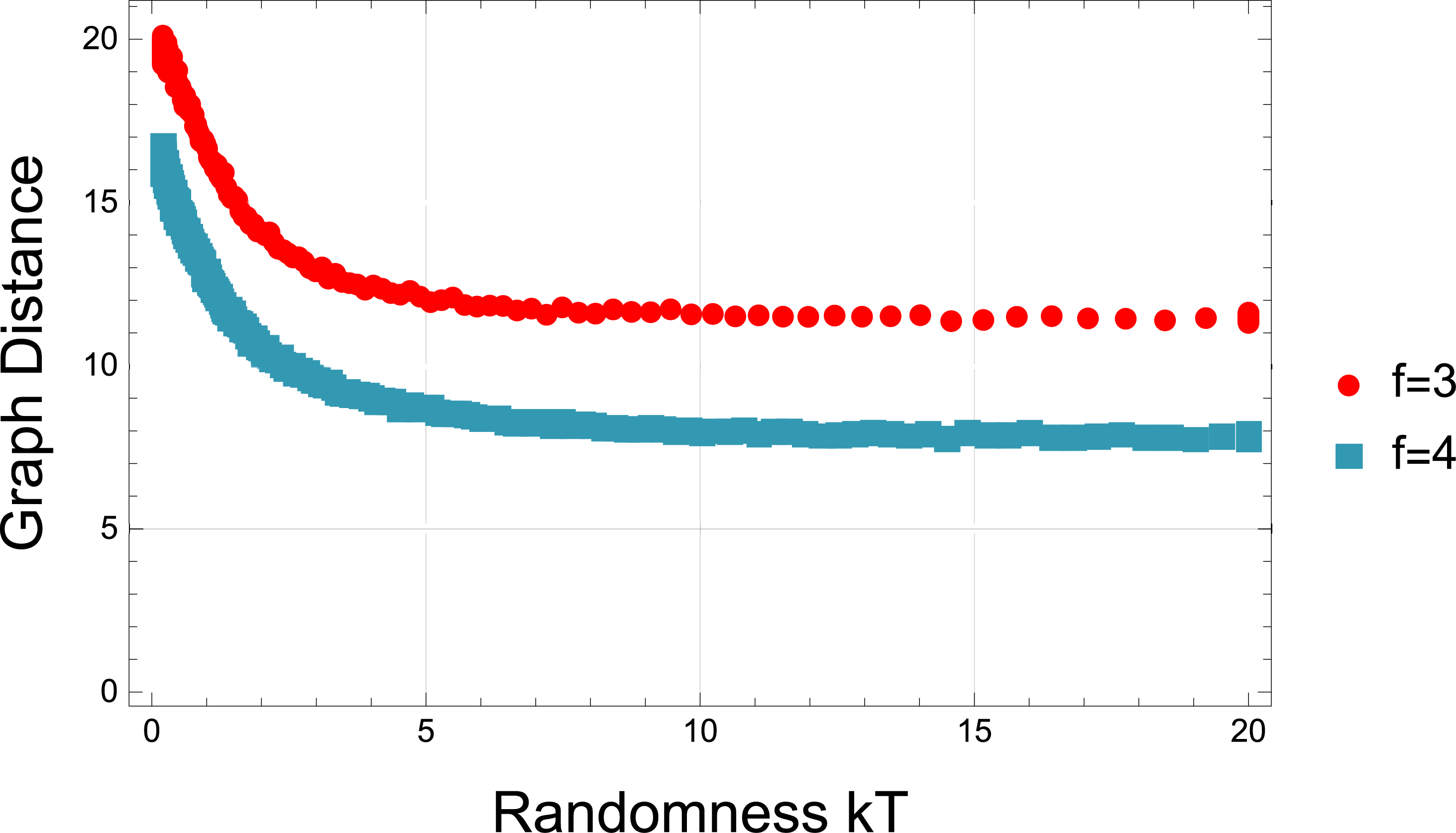}
\caption{The mean graph distances between randomly chosen pairs of vertices of regular graphs against the rate of randomness.}
\label{fig_GraphDistances}
\end{center}
\end{figure}

In figure \ref{fig_GraphDistances}, we show the dependence of mean graph distances on the rate of randomization.
The red circles corresponds to the data of $3$-regular graphs and the blue squares to those of $4$-regular graphs.
The mean graph distance decreases as $kT$ increases when $kT<5$.
They are constant if $kT>5$.
That convergence originates from the finite size of the graph:
if an uniform regular graph, the mean graph distance is proportional to $\log N$.
As it is lower limit of the mean graph distance, the level of the convergence depends on $N$.
For the target level of the rate of randomization $kT=0.20$, the mean graph distance of $3$-regular graphs was about $19.5$ and that of $4$-regular graphs was about $16$.
For the same rate of randomization, the mean graph distance of $3$-regular graphs is larger than that of $4$-regular graphs.

It should be remarkable that the relaxation of graph distances requires more switches than the relaxation of energies.

The rate of randomization $kT$ defines the mean graph distance between a randomly chosen pair of vertices.
Before randomization, the graph neighborhood up to distance $d$ has almost $f\times(f-1)^{d-1}$ vertices when $d$ is not too large since it has few cycles.
Therefore, the graph neighborhood up to distance $\log{N}/\log{(f-1)}$ has almost all vertices of the graph.
The maximum graph distance and the mean graph distance is proportional to $\log{N}/\log{(f-1)}$.
After randomization in low $kT$, connections of the graph are ordered by the geometrical coordinates.
The number of vertices in the graph neighborhood up to $d$ is proportional to the number of geometrically neighbor vertices $d^{\mathcal{D}}$. 
Here $\mathcal{D}$ is the dimension of the embedding space.
Therefore, the graph neighborhood up to distance $N^{1/\mathcal{D}}$ has almost all vertices of the graph.
The maximum graph distance and the mean graph distance is proportional to $N^{1/\mathcal{D}}$.

\begin{figure}
\begin{center}
\includegraphics[width=7.cm]{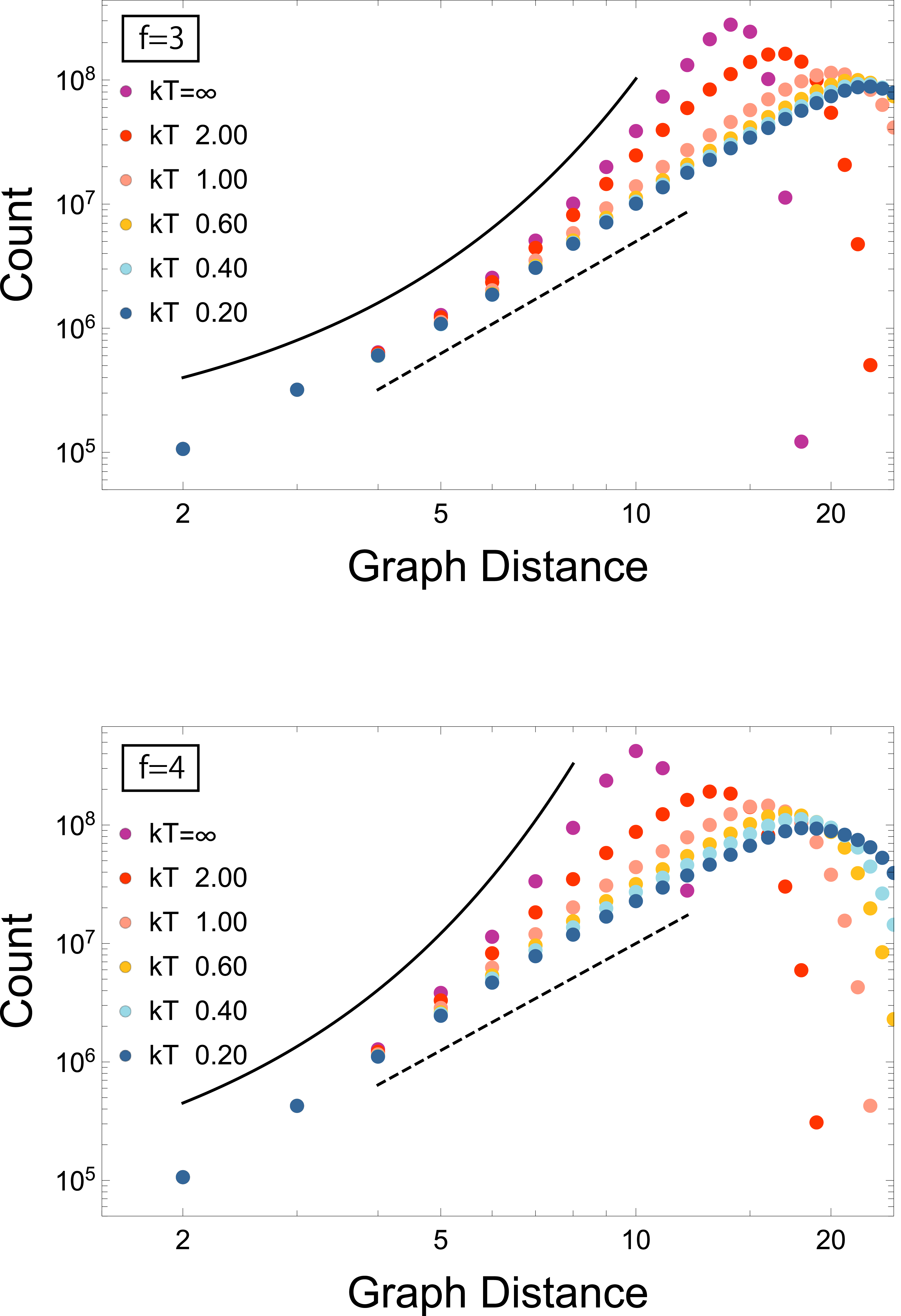}
\caption{The Double Logarithmic plot of the histogram of the graph distances between all pairs of vertices of regular graphs.}
\label{fig_GraphDistanceHistogram}
\end{center}
\end{figure}

In figure \ref{fig_GraphDistanceHistogram}, we generated $10$ regular graphs of $10648$ vertices randomized with a rate of randomness and show the number of vertex pairs against their graph distances $d$. 
The upper panel shows the plot for $3$-regular graphs and the bottom panel shows that for $4$-regular graphs.
For each panel, the purple data points represents the numbers of vertex pairs at distance $d$ of regular graphs before randomization, i.e. the uniform regular graphs.
The red data points represents those of regular graphs randomized in $kT=2.0$, the pink ones those in $kT=1.0$, the yellow ones those in $kT=0.6$, the pale blue ones those in $kT=0.4$ and the blue ones those in $kT=0.2$.
The solid lines are proportional to $(f-1)^d$ and the dashed lines are proportional to $d^3$.

The number of vertices at distance $d$ of uniform regular graphs increased exponentially as $d$ increased.
In contrast, that of regular graphs randomized in low $kT$ increased in proportion to $d^3$.

Good embeddings of a uniform regular graph are too concentrated as a polymer network, whose volume is proportional to $(\log{N})^3$
since the mean graph distance is proportional to $\log N$.
Here 'good' embeddings are embeddings where a graph distance and Euclidean distance between each pair of vertices are proportional.
It means a uniform regular graph is not suitable for the representation of a polymer network and a lattice network or a regular graph randomized in finite $kT$ is suitable.

\subsection{Mean centralities}

The table \ref{tab_MeanCent_f3} shows the mean closeness centralities of all vertices of randomized $3$-regular graphs 
and the table \ref{tab_MeanCent_f4} those of randomized $4$-regular graphs.

\begin{table}[htbp]
\centering
\begin{tabular}{cccc} \hline
 $kT$ & Mean & SD & Err \\ \hline
 $\infty$ & $0.0872511$ & $0.000343166$ & $1.05165\times 10^{-6}$  \\
 $2.00$ & $0.0693403$ & $0.00157994$ & $4.8418\times 10^{-6}$ \\
 $1.00$ & $0.0590339$ & $0.00143369$ & $4.3936\times 10^{-6}$ \\
 $0.60$ & $0.0540534$ & $0.00123554$ & $3.7864\times 10^{-6}$ \\
 $0.40$ & $0.0520149$ & $0.00107014$ & $3.2795\times 10^{-6}$ \\
 $0.20$ & $0.050753$ & $0.000911186$ & $2.7924\times 10^{-6}$ \\ \hline
\end{tabular}
\caption{Centralities for $f=3$ regular graphs.}
\label{tab_MeanCent_f3}
\end{table}

\begin{table}[h]
\centering
\begin{tabular}{cccc} \hline
 $kT$ & Mean & SD & Err \\ \hline
 $\infty$ & $0.128453$ & $0.000349456$ & $1.07092\times 10^{-6}$ \\
 $2.00$ & $0.0940562$ & $0.00179894$ & $5.51294\times 10^{-6}$ \\
 $1.00$ & $0.0781298$ & $0.00139067$ & $4.26177\times 10^{-6}$ \\
 $0.60$ & $0.0703325$ & $0.00108802$ & $3.33428\times 10^{-6}$ \\
 $0.40$ & $0.0663161$ & $0.000855799$ & $2.62264\times 10^{-6}$ \\
 $0.20$ & $0.061115$ & $0.000517275$ & $1.58521\times 10^{-6}$ \\ \hline
\end{tabular}
\caption{Centralities for $f=4$ regular graphs.}
\label{tab_MeanCent_f4}
\end{table}

The closeness centrality of the vertex $i$ is given  by the inverse of the average graph distance from $i$ to all other vertices.
The mean graph distance depended on the curves shown in \ref{fig_GraphDistanceHistogram}.
Thus the mean centrality decreased as the rate of randomization $kT$ increased.
Remark it will converge to zero as the number of all vertices $N$ increases since the expected graph distance $d$ glows as $N$ glows.

\subsection{Variance of Resistance Distances and Centralities}

While the mean of the resistance distances did not depend on the rates of randomness $kT$, the standard deviations depended.
Let us consider the origin of the infinite standard deviations.
It may be occurred from some inhomogeneity of the regular graphs.

\begin{figure}[htb]
\begin{center}
\includegraphics[width=6cm]{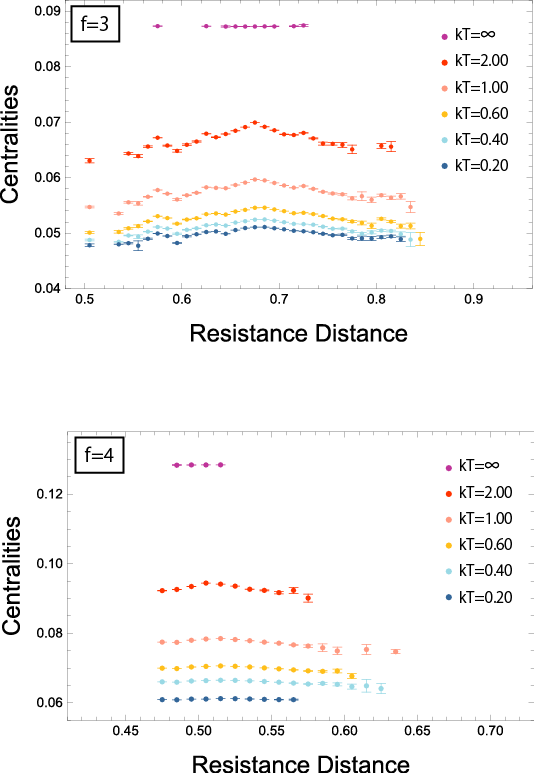}
\caption{Closeness centralities against resistance distances.}
\label{fig_ResistanceCents}
\end{center}
\end{figure}

The figure \ref{fig_ResistanceCents} shows the (in)dependence of the centrality on the resistance distances:
The purple points correspond to the data of regular graphs before randomization for each panel.
The red data points to those of regular graphs randomized in $kT=2.0$, the pink ones to those of $kT=1.0$, the yellow ones to those of $kT=0.6$, the pale blue ones to those of $kT=0.4$ and the blue ones to those of $kT=0.2$.
The y coordinates show the closeness centrality of a vertex and the x coordinates show the resistance distance from the vertex to a neighbor vertex. 
As $kT$ decreased, centralities decreased monotonically, but the standard deviation of the resistance distances did not.
Furthermore, centralities seems not to depend on the resistance distances.

We should conclude the closeness centrality isn't effective when we discuss the deviation of the resistance distances of the homogeneous networks.
The closeness centrality includes the information about all graph distances from a vertex to all other vertices, it should be too global.
The resistance distance depended on the local structure such as a graph neighborhood up to small $d$.

\end{document}